\documentclass[11pt,a4paper]{article}
\usepackage{authblk}
\usepackage{lmodern}
\usepackage{jheppub}
\usepackage{amsmath}
\usepackage{mathtools}
\usepackage{tikz}
\usepackage{empheq}
\usepackage{enumitem}
\usepackage{graphics}
\usepackage{wrapfig}
\usepackage{caption}
\usepackage{natbib}
\usepackage{xcolor}
\usepackage{framed}
\usepackage{array}
\usepackage{dsfont}
\definecolor{shadecolor}{gray}{0.925}

\usepackage[cbgreek]{textgreek}

\def\sideremark#1{\ifvmode\leavevmode\fi\vadjust{\vbox to0pt{\vss
 \hbox to 0pt{\hskip\hsize\hskip1em
 \vbox{\hsize3cm\tiny\raggedright\pretolerance10000
 \noindent #1\hfill}\hss}\vbox to8pt{\vfil}\vss}}}%

\newcommand{\bi}{\begin{itemize}}
\newcommand{\ei}{\end{itemize}}
\newcommand{\bea}{\begin{align}}
\newcommand{\eea}{\end{align}}
\newcommand{\be}{\begin{equation}}
\newcommand{\ee}{\end{equation}}

\makeatletter
\renewcommand*\env@matrix[1][\arraystretch]{%
  \edef\arraystretch{#1}%
  \hskip -\arraycolsep
  \let\@ifnextchar\new@ifnextchar
  \array{*\c@MaxMatrixCols c}}
\makeatother

\author[\ensuremath{a,b}]{Lorenzo IACOBACCI}
\author[\ensuremath{c}]{Charlotte SLEIGHT}
\author[\ensuremath{a},\ensuremath{b},\ensuremath{d}]{\quad Massimo TARONNA}

\affiliation[\ensuremath{a}]{Dipartimento di Fisica ``Ettore Pancini'', Universit\`a degli Studi di Napoli Federico II, \\Monte S. Angelo, Via Cintia, 80126 Napoli, Italy}

\affiliation[\ensuremath{b}]{INFN, Sezione di Napoli, Monte S. Angelo, Via Cintia, 80126 Napoli, Italy}

\affiliation[\ensuremath{c}]{Centre for Particle Theory and Department of Mathematical Sciences, \\ Durham University, Durham, DH1 3LE, U.K.}

\affiliation[\ensuremath{d}]{Scuola Superiore Meridionale, Universit\`a degli Studi di Napoli Federico II,\\ Largo San Marcellino 10, 80138 Napoli, Italy}

\emailAdd{lorenzo.iacobacci@unina.it,charlotte.sleight@durham.ac.uk, massimo.taronna@unina.it}

\title{\centering \huge From Celestial Correlators to AdS, and back}

\abstract{We present a general relation between celestial correlation functions in $d$-dimensions and Witten diagrams in $\left(d+1\right)$-dimensional Euclidean anti-de Sitter (EAdS) space, to all orders in perturbation theory. Contact diagram processes are proportional to contact Witten diagrams and particle exchanges can be recast as a continuum of particle exchanges in EAdS where the exchanged particles carrying unitary Principal Series representations of $SO\left(d+1,1\right)$. One can then try to import familiar EAdS techniques to study the properties of celestial correlators. In this work we use this relation to infer the analytic structure of the spectral density in the conformal partial wave expansion of celestial correlators which, at least perturbatively, should be a meromorphic function of the spectral parameter. We also discuss non-perturbative constraints from unitarity in Euclidean Conformal Field Theory, which requires positivity of the spectral density. This extends similar relations recently uncovered between boundary correlation functions in de Sitter space and Witten diagrams in EAdS, suggesting that EAdS could play a central role in efforts towards holography for all lambdas.}

\begin{document}

\begin{flushright}    
\texttt{}
\end{flushright}

\maketitle

\newpage

\section{Introduction}\label{sec::Intro}

Scattering amplitudes are fundamental observables in Minkowski space. There are issues however in extending scattering amplitudes to curved backgrounds, which are owing to difficulties in defining asymptotic states. In anti-de Sitter (AdS) and de Sitter (dS) space typically one considers correlation functions extrapolated to the boundaries at infinity. This lies at the heart of the AdS/CFT correspondence and has led to a new understanding of scattering in AdS space, as well as observables in dS space and cosmology. It would therefore be natural to study such types of observables also in cases where the usual notion of S-matrix makes sense. In this way one might be able to shed light on the properties of such new types of observables from the known properties of S-matrix amplitudes and as well put observables defined on different space-times on a similar footing. Mathematically, when both S-matrix and boundary correlators can be defined, going from one to the other is equivalent to acting with intertwiner operators that recast the corresponding observables in terms of a different basis of operators.

\vskip 4pt
Along these lines, in \cite{deBoer:2003vf,Cheung:2016iub,Pasterski:2016qvg,Pasterski:2017kqt} it has been shown that scattering amplitudes can be recast as $d$-dimensional Euclidean conformal field theory correlators by expressing them in a basis of $SO\left(d+1,1\right)$-primary solutions to the free equations of motion. Such correlators naturally live on the celestial sphere at null infinity, where the Lorentz group acts as the $d$-dimensional Euclidean conformal group $SO\left(d+1,1\right)$. Celestial holography (for reviews see e.g. \cite{Raclariu:2021zjz,Pasterski:2021rjz,McLoughlin:2022ljp,Pasterski:2021raf}) then postulates that there is a dual Conformal Field Theory (CFT) description. Note that these would not be the standard correlators in Euclidean CFT are encountered as Wick rotations of correlators in Lorentzian CFT, and for this reason we distinguish such correlators as \emph{celestial} correlators. They are defined on the celestial sphere at null infinity, which lacks a standard notion of locality and time, and are therefore not necessarily bound to satisfy the usual Osterwalder-Schrader axioms. Efforts to understand the properties of celestial correlators so far include the study of symmetries and translating of properties of scattering amplitudes in the standard plane-wave basis to the conformal basis, see e.g. \cite{Kapec:2014opa,Kapec:2016jld,Cheung:2016iub,Lam:2017ofc,Stieberger:2018edy,Stieberger:2018onx,Fan:2019emx,Nandan:2019jas,Pate:2019mfs,Adamo:2019ipt,Puhm:2019zbl,Guevara:2019ypd,Law:2019glh,Fotopoulos:2019vac,Banerjee:2020kaa,Fan:2020xjj,Donnay:2020guq,Fotopoulos:2020bqj,Casali:2020vuy,Banerjee:2020zlg,Banerjee:2020vnt,Pasterski:2020pdk,Pasterski:2021fjn,Jiang:2021ovh,Kapec:2022axw,Guevara:2021abz,Strominger:2021lvk,Himwich:2021dau,Donnay:2022sdg,Pasterski:2022djr}. 

\vskip 4pt
This is to be contrasted with the situation in anti-de Sitter space, where the isometry group of $\left(d+1\right)$-dimensional AdS acts as the $d$-dimensional Lorentzian conformal group $SO\left(d,2\right)$ on the boundary at spatial infinity. In this case the conformal boundary has a time direction, meaning that unitarity and causality in AdS space is intimately tied to unitarity and causality of the boundary system -- which is a standard Lorentzian CFT defined non-perturbatively by conformal symmetry, unitarity and an associative (and, in particular, convergent) operator product expansion. It is natural to ask if any of our understanding from the AdS case can be adapted to celestial holography. This is further motivated from the simple observation that celestial and AdS boundary correlators can be placed on a similar footing by Wick rotating the AdS to Euclidean AdS, so that they satisfy the same conformal Ward identities. Any differences in the way they encode consistent physics therefore lies in the freedom left over after taking constraints from conformal Ward identities into account. 

\vskip 4pt
It is worth noting that, almost in parallel, very similar questions are being posed in the context of de Sitter (dS) space. Indeed, the isometry group of $\left(d+1\right)$-dimensional dS acts as the $d$-dimensional Euclidean conformal group $SO\left(d+1,1\right)$ on the boundaries at past and future infinity, with boundary correlation functions constrained by the same conformal Ward identities as their celestial and Euclidean AdS counterparts. Much like their celestial counterparts above, such correlators are not necessarily subject to the usual Osterwalder-Schrader axioms since the dS boundary is purely spatial -- which obscures how they arise from a consistent picture of unitary time evolution in the bulk of dS. Efforts that aim to understand the properties of dS boundary correlators and the corresponding constraints on the physics have come to be known as the ``Cosmological Bootstrap" (see \cite{Baumann:2022jpr} for a recent review), which has drawn inspiration from both the S-matrix and Conformal Bootstrap programmes. 

\begin{figure}[t]
    \centering
    \includegraphics[width=0.75\textwidth]{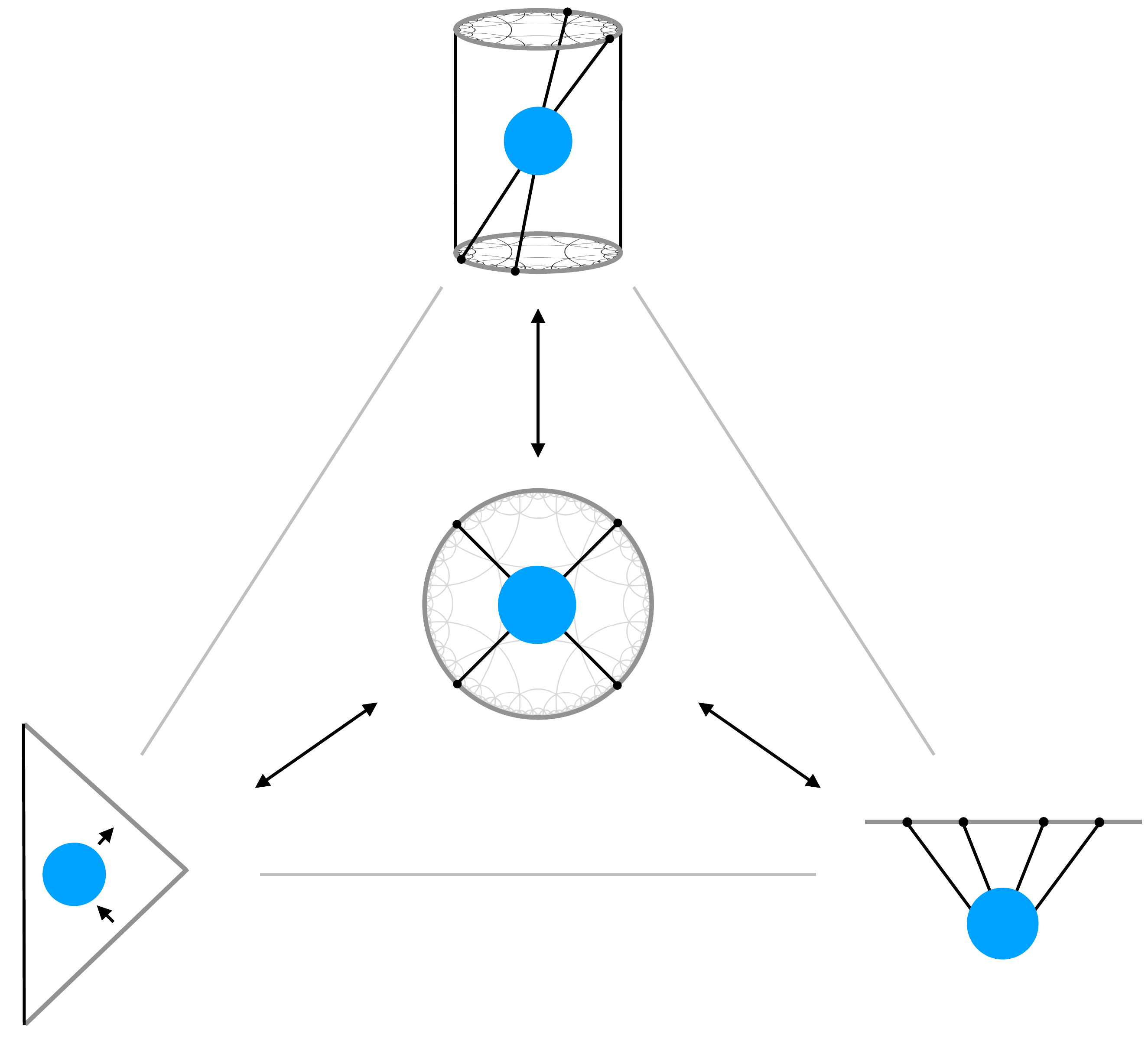}
    \caption{The Holographic Triangle: A cartoon representation of how observables at infinity in different maximally symmetric spaces can all be recast in terms of boundary observables in EAdS. Top: $\Lambda <0$, bottom-right: $\Lambda >0$ and bottom-left: $\Lambda=0$. The same theory on different spaces will analytically continue to different theories on EAdS. We stress that the analytic continuation is a only proper Wick rotation between AdS and EAdS.}
    \label{fig::holotri}
\end{figure}

\vskip 4pt
Among these results, it has been shown \cite{Sleight:2020obc} that any perturbative contribution to dS boundary correlators in the Bunch-Davies (Euclidean) vacuum can be recast as a linear combination of corresponding Witten diagrams in Euclidean AdS --- generated by the same collection of particles and couplings as the original process in dS space. This opens up the possibility to import the wealth of techniques, results and understanding from AdS to dS, with the caveat that one might have to consider a non-unitary process in AdS to study dS physics since the unitarity irreducible representations of the dS and AdS isometry groups (respectively, $SO\left(d+1,1\right)$ and $SO\left(d,2\right)$) do not coincide (though they are overlapping). Such identities between boundary correlators in dS and AdS were first explicitly constructed at all orders in perturbation theory by recasting Schwinger-Keldysh propagators in dS as certain analytic continuations of propagators for the same particle in EAdS \cite{Sleight:2019hfp,Sleight:2020obc}. More recently, however, they have been shown to follow from certain basic consistency criteria \cite{Sleight:2021plv,Baumann:2022jpr}. In particular, given that processes in dS and EAdS satisfy the same conformal Ward identities, the relations between them follow by requiring single-valuedness as a function of the complex cross-ratios and consistent on-shell factorisation. dS boundary correlators are only single-valued in the Bunch-Davies vacuum of dS \cite{Chen:2006nt,Holman:2007na,LopezNacir:2011kk,Flauger:2013hra,Aravind:2013lra} and single-valuedness is an important property of AdS boundary correlators in the Euclidean region following from the fact that singularities should only appear when one crosses light-cones. 

\vskip 4pt
Notice that one could run the same argument for celestial correlators under the assumption that they are also single-valued functions of the cross ratios. In this work we show that this is indeed the case.\footnote{See \cite{Lam:2017ofc,Casali:2022fro,deGioia:2022nkq} for relationships of a different kind between celestial correlators and Witten diagrams.} This draws inspiration from the idea \cite{deBoer:2003vf} that $(d+2)$-dimensional Minkowski space $\mathbb{M}^{d+2}$ is foliated by constant curvature hypersurfaces, which are either dS$_{d+1}$ or EAdS$_{d+1}$ depending on whether one is inside or outside the light-cone, and then apply holography to each slice. We show that the conformal primary wavefunction in each region of $\mathbb{M}^{d+2}$ can be expressed as an analytic continuation of the corresponding bulk-to-boundary propagator in EAdS$_{d+1}$ (i.e. of the same scaling dimension), times a function encoding the radial dependence. This allows us to establish that, like in dS, contact diagram contributions to celestial correlators in $d$-dimensions are proportional to their corresponding contact Witten diagrams in EAdS$_{d+1}$, and we give the proportionality constant explicitly. This confirms that such contact contributions to celestial correlators are single-valued solutions to the conformal ward identities. For processes involving particle exchanges, one uses that a field on $\mathbb{M}^{d+2}$ can be decomposed in terms of a continuum of fields carrying Principal Series representations of $SO\left(d+1,1\right)$ that live on the dS$_{d+1}$ or EAdS$_{d+1}$ hypersurfaces \cite{deBoer:2003vf}. Conformal Ward identities and single-valuedness then imply that an exchange contribution to a celestial correlator can be decomposed in terms of a continuum of corresponding exchange Witten diagrams in EAdS$_{d+1}$, with the exchanged particles carrying unitary Principal Series representations. The coefficient of each individual exchange Witten diagram is fixed by on-shell factorisation, meaning that it is given by the product of coefficients that convert each contact subdiagram to their celestial counterpart.

\vskip 4pt
We briefly explore some consequences of the relation between perturbative celestial correlators and Witten diagrams in EAdS. In particular, it implies that, like their AdS counterparts, celestial correlators admit a conformal partial wave expansion with meromorphic spectral density (at least perturbatively). One can then obtain their expansion into Conformal Blocks which, combined with crossing symmetry, lies at the centre of the bootstrap of standard Lorentzian CFTs. We also discuss (non-perturbative) constraints from unitarity in Euclidean conformal field theory at the level of the conformal partial wave expansion, which translates into positivity of the spectral density.

\vskip 4pt
Let us conclude the introduction by noting that these results suggest that Euclidean anti-de Sitter space could play a central role in unravelling holography on maximally symmetric spaces for all $\Lambda$. See figure \ref{fig::holotri}. Indeed, often one defines a theory via its Wick rotation to Euclidean space. For $\Lambda <0$, the Wick rotation from Lorentzian AdS to Euclidean AdS defines the Lorentzian theory by the usual Osterwalder-Schrader axioms. For $\Lambda > 0$, standard Wick rotation at the level of the embedding Minkowski space would define the theory on the sphere. The above results suggest however that in the Bunch-Davies vacuum one can alternatively define dS boundary observables (at least perturbatively) as boundary observables in EAdS that satisfy certain (non-standard) criteria that would ensure their consistency as a theory in dS. The theory in dS and its counterpart in EAdS are related by an analytic continuation given in \cite{Sleight:2020obc}. The results of this paper suggest a similar picture for celestial correlators, which perturbatively can be recast as boundary observables in EAdS that satisfy other (non-standard) criteria that would ensure their consistency as a theory in Minkowski space. In this case celestial correlators and boundary correlators in EAdS are related by analytic continuation and integrating over the dependence on the curvature radius.

\paragraph{Outline.} This paper is organised as follows. In section \ref{sec::Hyperslice} we review the hyperbolic slicing of Minkowski space, the conformal boundary, and the radial reduction of scalar fields onto each slice -- which is implemented via the Kantorovich-Lebedev transform. In section \ref{sec::CPB} we review the conformal primary basis for scattering amplitudes, introducing conformal primary wave functions and celestial correlators. In section \ref{sec::CCasWD}, implementing the hyperbolic slicing of $\mathbb{M}^{d+2}$, we show that the conformal primary wavefunction in each region of Minkowski space can be written as a bulk-to-boundary propagator in EAdS$_{d+1}$ (appropriately analytically continued) with the same scaling dimension, times a function encoding the radial dependence. From this it follows in section \ref{subsec::contact} that contact diagram contributions to celestial correlators are proportional to their contact Witten diagram counterparts in EAdS$_{d+1}$, with the proportionality constant arising from integrating out the dependence on the curvature radius. We compare with some known results for contact diagrams in celestial holography and observe a slight discrepancy with results that obtain celestial correlators from momentum space scattering amplitudes, which seems to arise from the fact that identities relating momentum space scattering amplitudes and celestial correlators entail commuting the divergent integral over the momentum. In section \ref{subsec::allorders} the perturbative relation between celestial correlators and Witten diagrams is extended to include processes that involve particle exchanges. In section \ref{subsec::spectralrep} these results are used to infer properties of the conformal partial wave expansion of celestial correlators, in particular meromorphicity of the spectral function, and non-perturbative constraints from unitarity.

\section{Hyperbolic slicing of Minkowski space}
\label{sec::Hyperslice}

\begin{figure}[htb]
    \centering
    \includegraphics[width=0.6\textwidth]{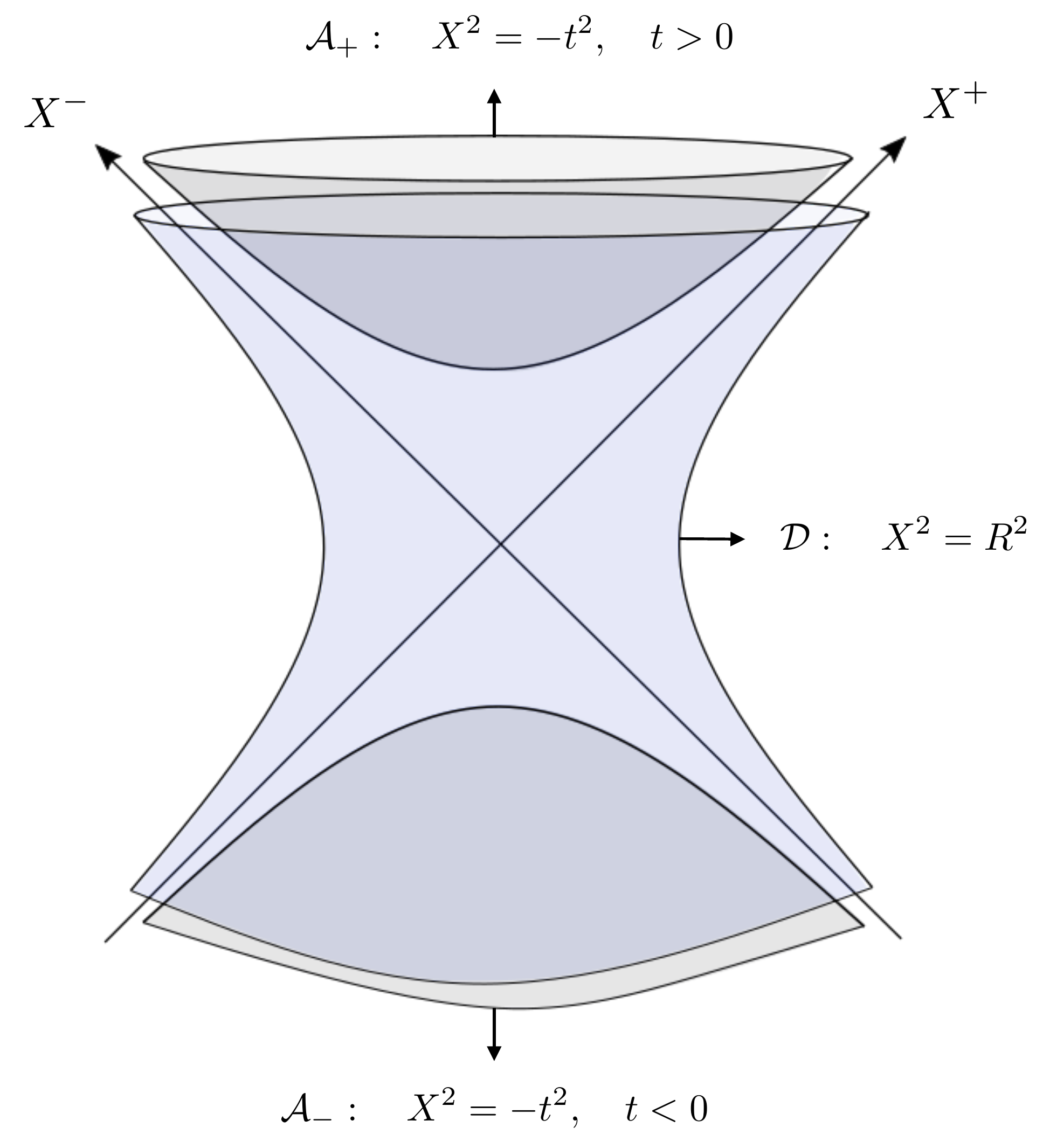}
    \caption{Hyperbolic slicing of Minkowski space. Region ${\cal D}$ outside the light-cone is foliated by co-dimension one de Sitter spaces and regions ${\cal A}_\pm$ inside the light-cone by the upper sheet (in region ${\cal A}_+$) or lower sheet (in region ${\cal A}_-$) of co-dimension one Euclidean anti-de Sitter spaces.}
    \label{fig::HMink}
\end{figure}

Consider $\left(d+2\right)$-dimensional Minkowski space $\mathbb{M}^{d+2}$ with Cartesian coordinates $X^M$, $M=0,\ldots, d+1$ and metric
\begin{equation}
    ds^2=-\left(dX^0\right)^2+\left(dX^1\right)^2+\ldots+\left(dX^{d+1}\right)^2. 
\end{equation} 
The light-cone, defined as 
\begin{equation}\label{LC}
     X^2=-\left(X^0\right)^2+\left(X^1\right)^2+\ldots+\left(X^{d+1}\right)^2=0,\end{equation}
naturally divides $\mathbb{M}^{d+2}$ into three regions:
\begin{subequations}
\begin{align}
    \mathcal{D}:&\quad X^2>0\,,\\
    \mathcal{A}_+:&\quad X^2<0\,,\qquad X_0>0,\\
    \mathcal{A}_-:&\quad X^2<0\,,\qquad X_0<0.
\end{align}
\end{subequations}
 Each region can be foliated with surfaces of constant curvature reflecting the $SO\left(d+1,1\right)$ symmetry. See figure \ref{fig::HMink}. For $\mathcal{A}_\pm$ these are $\left(d+1\right)$-dimensional Euclidean anti-de Sitter spaces with constant radius $t$:
\begin{equation}
    X^2=-t^2. \label{EAdSt}
\end{equation}
A natural set of coordinates for this foliation of ${\cal A}_\pm$ is
\begin{subequations}\label{EAdScoords}
\begin{align}
    \mathcal{A}_+:&\quad X^M=t(\cosh y,\sinh y\, \hat{n}),\quad t>0,\quad y>0,\\
    \mathcal{A}_-:&\quad X^M=t(\cosh y,\sinh y\, \hat{n}),\quad t<0,\quad y<0, 
\end{align}
\end{subequations}
which is the analogue of spherical coordinates. ${\hat n}$ is a $d$-dimensional unit vector ${\hat n}\cdot {\hat n}=1$ parameterising the $d$-dimensional unit sphere. For ${\cal D}$ instead, the foliating surfaces are $\left(d+1\right)$-dimensional de Sitter space-times with radius $R$:
\begin{equation}
    X^2=R^2. \label{dSr}
\end{equation}
In this case, the natural set of coordinates are
\begin{equation}\label{dScoords}
    \mathcal{D}:\quad X^M=R(\sinh\tau,\cosh\tau\,\hat{n}),
\end{equation}
where $\tau\in(-\infty,\infty)$ and ${\hat n}$ parameterise the $d$-dimensional unit sphere as before. It is convenient to divide $\mathcal{D}$ into two regions corresponding to expanding and contracting patches of the dS hypersurfaces \eqref{dSr} respectively:
\begin{subequations}
\begin{align}
    \mathcal{D}_+:&\quad X^2>0\,,\qquad X^+>0,\\
    \mathcal{D}_-:&\quad X^2>0\,,\qquad X^+<0,
\end{align}
\end{subequations}
with light cone coordinate $X^+=X^{0}+X^{d+1}$. The hyperbolic slicing of the individual regions ${\cal A}_\pm$ and ${\cal D}_\pm$ are naturally described by Poincar\'e coordinates. In fact, in the hyperbolic slicing $\mathbb{M}^{d+2}$ splits up into four Poincar\'e patches:
\begin{subequations}\label{poincoords}
\begin{align}
    \mathcal{A}_+:&\qquad X^M=+\frac{t}{z}\left(\frac{1+z^2+\vec{x}^2}{2},\frac{1-z^2-\vec{x}^2}{2},\vec{x}\right),& t&>0, & z&\in\left[0,\infty\right)\\
    \mathcal{A}_-:&\qquad X^M=+\frac{t}{z}\left(\frac{1+z^2+\vec{x}^2}{2},\frac{1-z^2-\vec{x}^2}{2},\vec{x}\right),& t&<0,& z&\in\left[0,\infty\right)\\
    \mathcal{D}_+:&\qquad X^M=-\frac{R}{\eta_+}\left(\frac{1-\eta^2_++\vec{x}^2}{2},\frac{1+\eta^2_+-\vec{x}^2}{2},\vec{x}\right),& R&>0,& \eta_+&\in\left[0,\infty\right)\\
    \mathcal{D}_-:&\qquad X^M=\frac{R}{\eta_-}\left(\frac{1-\eta^2_-+\vec{x}^2}{2},\frac{1+\eta^2_--\vec{x}^2}{2},\vec{x}\right),& R&>0,& \eta_-&\in\left[0,\infty\right),
\end{align}
\end{subequations}
where $\vec{x}\in \mathbb{R}^d$ parameterises the boundary directions.

\vskip 4pt
\paragraph{Conformal Boundary.} $\mathbb{M}^{d+2}$ has a conformal boundary at both past and future null infinity, which can be identified with the projective cone of light rays via
\begin{equation}\label{PrLC}
    Q^2=0, \qquad Q \equiv \lambda Q, \qquad \lambda \in \mathbb{R}_{>0}.
\end{equation}
Both the past and future conformal boundaries are $d$-dimensional spheres, which we shall denote by $S^-_d$ and $S^+_d$ respectively. This can be seen by defining new projective coordinates
\begin{align}
    \xi_1 = Q^1/Q^0, \quad \xi_2 = Q^2/Q^0, \quad \ldots \quad, \quad \xi_{d+1} = Q^{d+1}/Q^0,
\end{align}
so that 
\begin{equation}
    \xi^2_1+\ldots+\xi^2_{d+1}-1=0.
\end{equation}
The sphere $S^-_d$ corresponds to $Q^0<0$ and $S^+_d$ to $Q^0>0$. The conformal boundaries $S^\pm_d$ of $\mathbb{M}^{d+2}$ are also conformal boundaries of each of its hyperbolic slices \eqref{EAdSt} and \eqref{dSr}, which asymptote to the lightcone \eqref{PrLC}. The region ${\cal A}_+$ is foliated by the upper sheet of the hyperboloids \eqref{EAdSt}, which have conformal boundary $S^+_d$. Likewise, the region ${\cal A}_-$ is foliated by the lower sheet of the hyperboloids \eqref{EAdSt}, which have conformal boundary $S^-_d$. In both cases the boundaries $S^\pm_d$ are approached at spatial infinity $|y| \to \infty$ with boundary coordinates:
\begin{subequations}
\begin{align}
    Q_+&\sim \tfrac{1}{2}\,te^y (1,\hat{n})&  &\text{with}&  t>0,\\
    Q_-&\sim \tfrac{1}{2}\,te^{-y} (1,-\hat{n})&  &\text{with}& t<0.
\end{align}
\end{subequations}
In region ${\cal D}$ the foliating surfaces \eqref{dSr} are de Sitter space-times, which have two conformal boundaries: $S^+_d$ at future infinity and $S^+_d$ at past infinity. In the parameterisation \eqref{dScoords} these are obtained in the limits $\tau \to \pm \infty$ with boundary coordinates: 
\begin{align}
    Q_+&\sim \tfrac12 R e^\tau(1,\hat{n}),& Q_- &\sim \tfrac12 R e^{-\tau}(-1,\hat{n}).
\end{align}

Since the $Q_\pm$ lie on the projective null cone \eqref{PrLC}, they can be re-scaled to the following canonical form which we employ throughout:
\begin{align}
    Q_+&\sim (1,\hat{n}),& Q_- &\sim (-1,\hat{n}).
\end{align}

\vskip 4pt
\paragraph{Scalar fields.} Consider now a scalar field $\phi\left(X\right)$ in $\mathbb{M}^{d+2}$ of mass $m$ with free equation of motion
\begin{equation}\label{KGeq}
    \left(\frac{\partial}{\partial X^M}\frac{\partial}{\partial X_M}-m^2\right)\phi\left(X\right)=0.
\end{equation}
The field $\phi\left(X\right)$ can be decomposed into fields that live on the hyperbolic slices \eqref{EAdSt} and \eqref{dSr} of $\mathbb{M}^{d+2}$ with unit radius by applying the Kantorovich-Lebedev transform \cite{deBoer:2003vf} (see appendix \ref{sec::KLT}). Focusing for concreteness on region ${\cal D}$ (analogous expressions hold for the other regions ${\cal A}_\pm$) this reads 
\begin{equation}\label{flatAdSdecomp}
    \phi\left(X\right) = \frac{1}{2}\int^{\frac{d}{2}+i\infty}_{\frac{d}{2}-i\infty}\frac{d\Delta}{2\pi i}\, \phi_{\Delta}({\hat X}) {\tilde K}_{\Delta-\tfrac{d}{2}}\left( mR\right),
\end{equation}
where
\begin{equation}\label{tildeK}
    {\tilde K}_{\Delta-\tfrac{d}{2}}\left(m R\right) = \frac{2R^{-d/2}}{\Gamma(\Delta-\tfrac{d}{2})}\,K_{\Delta-\tfrac{d}{2}}(mR),
\end{equation}
which is proportional to a modified Bessel function of the second kind, $K_{\Delta-\tfrac{d}{2}}(mR)$. The field $\phi_{\Delta}({\hat X})$ satisfies the massive field equation on the de Sitter hyperboloid \eqref{dSr} with $R=1$
\begin{equation}
    \left(\nabla^2_{\text{dS}}-\Delta\left(d-\Delta\right)\right)\phi_{\Delta} = 0.
\end{equation}
What we see is that the radial reduction of a field $\phi\left(X\right)$ in $\mathbb{M}^{d+2}$ onto the hyperbolic slices \eqref{EAdSt} and \eqref{dSr} yields an infinite number of fields on EAdS$_{d+1}$ and dS$_{d+1}$, respectively, which carry principal series representations $\Delta \in \frac{d}{2}+i\mathbb{R}$ of $SO\left(d+1,1\right)$.\footnote{Much like in the standard CFT literature \cite{Dobrev:1975ru,Dobrev:1977qv,Mack:2009mi,Costa:2012cb,Caron-Huot:2017vep}, the $\Delta \in \frac{d}{2}+i\mathbb{R}$ on the Principal Series can be analytically continued to $\Delta \in \mathbb{C}$ \cite{Donnay:2020guq}. In the context of celestial correlators this is relevant for the treatment of conformally soft Goldstone modes \cite{Pasterski:2017kqt,Cheung:2016iub,Donnay:2018neh,Fan:2019emx,Pate:2019mfs,Adamo:2019ipt,Puhm:2019zbl,Guevara:2019ypd}.} Applying holography to each slice, the existence of a dual conformal field theory description of $\mathbb{M}^{d+2}$ living on the boundary spheres $S^\pm_d$ was postulated in \cite{deBoer:2003vf}. Upon appropriate anti-podal identification of $S^+_d$ and $S^-_d$ defined in the following section, the correlation functions of the corresponding dual conformal operators have come to be known as \emph{celestial correlators}. These correspond to scattering amplitudes evaluated in a conformal basis \cite{Cheung:2016iub,Pasterski:2016qvg,Pasterski:2017kqt} and are reviewed in the following section.

\section{Conformal primary basis for scattering amplitudes}
\label{sec::CPB}

 Particles of mass $m$ and definite momentum $p$ in $\mathbb{M}^{d+2}$ correspond to the planewave solutions $e^{i p \cdot X}$ to \eqref{KGeq} where ${\hat p} =p/m$ parameterises the $\left(d+1\right)$-dimensional two-sheeted hyperboloid:
\begin{equation}\label{MS}
    H_{d+1}: \quad {\hat p}^2=-1.
\end{equation}
Another basis of delta function normalisable solutions to \eqref{KGeq} is given by what are known as \emph{conformal primary wavefunctions} \cite{Pasterski:2016qvg,Pasterski:2017kqt}. These are fields $\phi_{\Delta}\left(X;Q\right)$ labelled by a conformal dimension $\Delta$ and boundary point $Q$ that transform as a conformal primary scalar operator in $d$-dimensions under $SO\left(d+1,1\right)$. To wit,
\begin{align}\label{CPWdef}
    \phi_{\Delta}\left(X;\lambda Q\right) = \lambda^{-\Delta}\phi_{\Delta}\left(X;Q\right),
\end{align}
where we recall that when parameterising the conformal boundary with the projective lightcone \eqref{PrLC} the $SO\left(d+1,1\right)$ transformations act linearly \cite{Costa:2011mg}. Let us first consider the Fourier transform $\phi_{\Delta}\left({\hat p};Q\right)$. Given that the momentum ${\hat p}$ lies on $H_{d+1}$, the condition \eqref{CPWdef} implies that is given by the bulk-to-boundary propagator $G_{\Delta}\left({\hat p};Q\right)$ for a scalar field on $H_{d+1}$ with conformal dimension $\Delta$,
\begin{equation}
    \phi_{\Delta}\left({\hat p};Q\right) \propto G_{\Delta}\left({\hat p};Q\right),
\end{equation}
where (see e.g. \cite{Costa:2014kfa})
\begin{subequations}\label{buboEAdS}
\begin{align}
    G_{\Delta}\left({\hat p};Q\right) &= \frac{C^{\text{AdS}}_{\Delta}}{\left(-2{\hat p}\cdot Q\right)^\Delta},\\
    C^{\text{AdS}}_{\Delta} &= \frac{\Gamma\left(\Delta\right)}{2\pi^{\frac{d}{2}}\Gamma\left(\Delta-\frac{d}{2}+1\right)}.
\end{align}
\end{subequations}
Conformal primary wavefunctions therefore have the Fourier expansion \cite{Pasterski:2016qvg,Pasterski:2017kqt}
\begin{equation}\label{PWtoCPWF}
    \phi_{\Delta}\left(X;Q\right)=\mathcal{N}_\Delta\left(\frac{m}{2\pi}\right)^{d/2}\int_{H_{d+1}}\left[d{\hat p}\right] G_{\Delta}\left({\hat p};Q\right)e^{i m {\hat p}\cdot X},
\end{equation}
with
\begin{align}
    \mathcal{N}_\Delta=\frac{\Gamma(\Delta-\tfrac{d}{2}+1)}{\sqrt{\pi}}\,.
\end{align}
Incoming ($-$) or outgoing ($+$) wavefunctions can be written down by constraining ${\hat p}$ to lie, say, on the upper sheet $H^+_{d+1}$ of the hyperboloid \eqref{MS}:\footnote{Note that we have normalised the conformal primary wave functions according to the canonical Klein-Gordon inner product:
\begin{align}\label{cKG}
    (\phi^{\pm}_{\Delta_1},\phi^{\pm}_{\Delta_2})=\pm2\pi i\,\delta(\Delta_1-\Delta_2)\,,
\end{align}
which differs from the normalisation used in e.g. \cite{Pasterski:2016qvg,Pasterski:2017kqt}.
}
\begin{equation}
    \phi^\pm_{\Delta}\left(X;Q_\pm\right)=\frac{\mathcal{N}_\Delta}{\Gamma(\Delta-\tfrac{d}{2})}\left(\frac{m}{2\pi}\right)^{d/2}\int_{H^+_{d+1}}\left[d{\hat p}\right] G_{\Delta}\left({\hat p};\pm Q_\pm\right)e^{\pm i m {\hat p}\cdot X},
\end{equation}
where $Q_\pm$ parameterises points on the boundary of $H^\pm_{d+1}$. Notice that a point $Q_-$ on the boundary of $H^-_{d+1}$ can be identified with a corresponding point $Q_+$ on the boundary of $H^+_{d+1}$ (and vice versa) via the antipodal identification:
\begin{equation}\label{apid}
    Q_+ := - Q_-.
\end{equation}
The incoming and outgoing wavefunctions can then be written as
\begin{equation}\label{CPWFintrep}
    \phi^\pm_{\Delta}\left(X;\pm Q\right)=\frac{\mathcal{N}_\Delta}{\Gamma(\Delta-\tfrac{d}{2})}\left(\frac{m}{2\pi}\right)^{d/2}\int_{H^+_{d+1}}\left[d{\hat p}\right] G_{\Delta}\left({\hat p};Q\right)e^{\pm i m {\hat p}\cdot X},
\end{equation}
where $Q$ is a boundary point on $H^+_{d+1}$. 

\vskip 4pt
Above we described a change of basis from plane waves to conformal primary wavefunctions
\begin{equation}
    e^{\pm i m {\hat p}\cdot X} \quad \leftrightarrow \quad  \phi^\pm_{\Delta}\left(X;\pm Q\right),
\end{equation}
which is implemented via \eqref{PWtoCPWF}. In the basis of conformal primary wavefunctions scattering amplitudes transform as correlation functions of conformal primary operators that, upon the anti-podal identification \eqref{apid}, are inserted on the $d$-dimensional sphere $S^+_d$ at the future conformal boundary of $\mathbb{M}^{d+2}$ which from this point on-wards we refer to as the \emph{celestial sphere}. The change of basis
\begin{equation}
    {\cal A}\left(\pm_1 p_1,\ldots,\pm_n\,p_n\right) \quad \leftrightarrow \quad  {\tilde {\cal A}}_{\Delta_1 \ldots \Delta_n}\left(\pm_1\,Q_1,\ldots,\pm_n\,Q_n\right),
\end{equation}
between an $n$-point scattering amplitude and an $n$-point celestial conformal correlation function is similarly implemented via \cite{Pasterski:2016qvg,Pasterski:2017kqt}
\begin{multline}\label{ccorr}
    {\tilde {\cal A}}_{\Delta_1 \ldots \Delta_n}\left(\pm_1\,Q_1,\ldots,\pm_n\,Q_n\right) =  \int_{H^+_{d+1}}\left[d{\hat p}_1\right] \ldots \left[d{\hat p}_n\right] \left(\,\prod\limits^n_{i=1}\frac{\mathcal{N}_{\Delta_i}}{\Gamma(\Delta_i-\tfrac{d}{2})}\left(\frac{m_i}{2\pi}\right)^{d/2}G_{\Delta_i}\left({\hat p}_i;Q_i\right)\,\right)   \\ \times {\cal A}\left(\pm_1\,p_1,\ldots, \pm_n\, p_n\right),
\end{multline}
where the notation $\pm_i$ refers to whether the $i$th external leg is incoming $\left(\pm_i=-\right)$ or outgoing $\left(\pm_i=+\right)$. 

\vskip 4pt
The change of basis \eqref{ccorr} has been employed in various works (see e.g. \cite{Raclariu:2021zjz,Pasterski:2021rjz,McLoughlin:2022ljp} for reviews) to translate features of scattering familiar from momentum space to the language of conformal correlators. On the other hand, given that celestial correlators transform as conformal correlators under $SO\left(d+1,1\right)$, it is natural to ask if they can be reformulated in terms of Witten diagrams in $H^+_{d+1}$, which we consider in the following section. Note that according to the definition of the conformal primary wave function, this would be the natural way to compute celestial correlators -- treating the conformal primary wave function as the analogue of the bulk-to-boundary propagator. In fact, in order to relate this picture to the momentum space S-matrix picture \eqref{ccorr}, one is required to commute the integral over Minkowski space with the divergent momentum integral. In the examples considered in section \ref{subsec::contact} we indeed seem to find a slight mismatch in the expressions computed in these two different ways.

\section{Celestial correlators as Witten diagrams}
\label{sec::CCasWD}

In this section we will show explicitly that a given perturbative celestial correlator can be expressed in terms of corresponding Witten diagrams on $H^+_{d+1}$. To this end, it is useful to evaluate the momentum integral in the representation \eqref{CPWFintrep} of conformal primary wave functions. This integral is only formal since it is divergent for real values of the mass $m$. One can however define the following convergent integrals:
\begin{subequations}\label{convintA}
\begin{align}
    \varphi^+_{\Delta}(X;Q_+)&=\frac{\mathcal{N}_\Delta}{\Gamma(\Delta-\tfrac{d}{2})}\left(\frac{m}{2\pi}\right)^{d/2}\int_{H^+_{d+1}}[d{\hat p}]\,G_{\Delta}\left({\hat p};Q_+\right)\,e^{ + m \,{\hat p}\cdot X}\,, \quad X \in {\cal A}_+,\\
    \varphi^-_{\Delta}(X;Q_-)&=\frac{\mathcal{N}_\Delta}{\Gamma(\Delta-\tfrac{d}{2})}\left(\frac{m}{2\pi}\right)^{d/2}\int_{H^-_{d+1}}[d{\hat p}]\,G_{\Delta}\left({\hat p};Q_-\right)\,e^{ + m \,{\hat p}\cdot X}\,, \quad X \in {\cal A}_-.
\end{align}
\end{subequations}
In the hyperbolic slicing we can write:
\begin{equation}\label{hyperfol0}
    X = R {\hat X}_{\text{AdS}},
\end{equation}
where in ${\cal A}_+$ we have $R > 0$ and in ${\cal A}_-$ we have $R < 0$, see \eqref{EAdScoords}. The integrals \eqref{convintA} are evaluated in Appendix \ref{sec::CPWint} and are given by the following closed form expression: 
\begin{subequations}\label{cpwconcf}
\begin{align}
    \varphi^+_{\Delta}(R,{\hat X}_{\text{AdS}};Q_+)&=\mathcal{N}_\Delta\, G_{\Delta}({\hat X}_{\text{AdS}};Q_+) \tilde{K}_{\Delta-\tfrac{d}2}(m R)\,, \\
    \varphi^-_{\Delta}(R,{\hat X}_{\text{AdS}};Q_-)&=\mathcal{N}_\Delta\, G_{\Delta}(-{\hat X}_{\text{AdS}};Q_-) \tilde{K}_{\Delta-\tfrac{d}2}(-m R)\,, 
\end{align}
\end{subequations}
which are factorised into the bulk-to-boundary propagator \eqref{buboEAdS} on $H^\pm_{d+1}$ and the function Bessel \eqref{tildeK} which, up to powers of $|R|$ and $m$, is the bulk-to-boundary propagator in Poincar\'e coordinates with the mass $m$ playing the role of the magnitude of the boundary momentum and $|R|$ the role of the Poincar\'e coordinate $z$.

\vskip 4pt
Closed form expressions for the conformal primary wavefunctions $\phi^\pm_{\Delta}$ over the whole of Minkowski space can be obtained from the closed form expression \eqref{cpwconcf} for $\varphi^\pm_{\Delta}$ by analytic continuation. By comparing with the integral expression \eqref{CPWFintrep} for $\phi^\pm_{\Delta}$, under such analytic continuations we require, for both $\varphi^\pm_{\Delta}$, that:
\begin{equation}\label{analx}
    X\to i X.
\end{equation}
At the level of the hyperbolic foliation \eqref{hyperfol0} this can be achieved for all regions ${\cal A}_\pm$ and ${\cal D}_\pm$ of $\mathbb{M}^{d+2}$ by Wick rotating $R$ and ${\hat X}_{\text{AdS}}$ respectively. These are summarised in the following.

\vskip 4pt
\paragraph{Outgoing modes.}

Let us first focus on the outgoing wavefunction $\phi^+_{\Delta}$. To obtain this in region ${\cal A}_\pm$ from $\varphi^+_{\Delta}\left(X;Q\right)$ we must stay on $H^+_{d+1}$ and Wick rotate $R$, which fixes:
\begin{subequations}\label{outapam}
\begin{align}
  &X \in {\cal A}_+: \qquad  \phi^+_{\Delta}(X;Q_+)=\varphi^+_{\Delta}(R, {\hat X}_{\text{AdS}};Q_+), \quad R = e^{+\frac{\pi i}{2}}t,\\
  &X \in {\cal A}_-: \qquad  \phi^+_{\Delta}(X;Q_+)=\varphi^+_{\Delta}(R,{\hat X}_{\text{AdS}};Q_+), \quad R = e^{-\frac{\pi i}{2}}|t|.
\end{align}
\end{subequations}
For the region ${\cal D}$, which is foliated by dS hypersurfaces \eqref{dSr}, we must Wick rotate ${\hat X}_{\text{AdS}}$ to the de Sitter hypersurface ${\hat X}^2_{\text{dS}}=1$. For the outgoing mode this fixes uniquely:
\begin{subequations}\label{outdpdm}
\begin{align}
  &X \in {\cal D}_+: \qquad  \phi^+_{\Delta}(X;Q_+)=\varphi^+_{\Delta}(R,{\hat X}_{\text{AdS}};Q_+), \quad {\hat X}_{\text{AdS}} =e^{+\frac{\pi i}{2}}{\hat X}_{\text{dS}},\\
  &X \in {\cal D}_-: \qquad  \phi^+_{\Delta}(X;Q_+)=\varphi^+_{\Delta}(R,{\hat X}_{\text{AdS}};Q_+), \quad {\hat X}_{\text{AdS}} =e^{-\frac{\pi i}{2}}\left(-{\hat X}_{\text{dS}}\right).
\end{align}
\end{subequations}

\vskip 4pt
\paragraph{Incoming modes.}

Similarly one obtain incoming wavefunctions. We summarise the results below:
\begin{subequations}\label{inall}
\begin{align}
  &X \in {\cal A}_+: \qquad  \phi^-_{\Delta}(X;Q_-)=\varphi^-_{\Delta}(R, {\hat X}_{\text{AdS}};Q_-), \quad R=e^{\frac{\pi i}{2}}t,\\
  &X \in {\cal A}_-: \qquad  \phi^-_{\Delta}(X;Q_-)=\varphi^-_{\Delta}(R,{\hat X}_{\text{AdS}};Q_-), \quad R=e^{-\frac{\pi i}{2}}|t|,\\
  &X \in {\cal D}_+: \qquad  \phi^-_{\Delta}(X;Q_-)=\varphi^-_{\Delta}(- R, {\hat X}_{\text{AdS}};Q_-), \quad {\hat X}_{\text{AdS}} =e^{-\frac{\pi i}{2}}{\hat X}_{\text{dS}},\\
  &X \in {\cal D}_-: \qquad  \phi^-_{\Delta}(X;Q_-)=\varphi^-_{\Delta}(- R, {\hat X}_{\text{AdS}};Q_-), \quad {\hat X}_{\text{AdS}} =e^{+\frac{\pi i}{2}}\left(-{\hat X}_{\text{dS}}\right).
\end{align}
\end{subequations}

\vskip 4pt
\paragraph{Conformal Primary Wave functions as EAdS propagators.} In \eqref{outapam}, \eqref{outdpdm} and \eqref{inall} above we identified the incoming and outgoing conformal primary wavefunctions $\phi^\pm_{\Delta}$ in $\mathbb{M}^{d+2}$ as analytic continuations of the functions $\varphi^\pm_{\Delta}$ defined in region ${\cal A}_\pm$. Inserting the closed-form expression \eqref{cpwconcf} for $\varphi^\pm_{\Delta}$, these can be summarised as follows:
\begin{subequations}\label{CPWFanals}
\begin{align}
& X \in {\cal A}_+: \quad   \phi^\pm_{\Delta}(X;Q_\pm) =  \mathcal{N}_\Delta G_{\Delta}(\pm{\hat X}_{\text{AdS}};Q_\pm) {\tilde K}_{\Delta-\frac{d}{2}}\left(m\, t e^{\pm \frac{\pi i}{2}}\right), \\
  & X \in {\cal A}_-: \quad   \phi^\pm_{\Delta}(X;Q_\pm) =  \mathcal{N}_\Delta G_{\Delta}(\pm{\hat X}_{\text{AdS}};Q_\pm) {\tilde K}_{\Delta-\tfrac{d}2}\left(m\, |t| e^{\mp \frac{\pi i}{2}}\right), \\
   &  X \in {\cal D}_+: \quad  \phi^\pm_{\Delta}(X;Q_\pm) =\mathcal{N}_\Delta G_{\Delta}(e^{+\frac{\pi i}{2}}{\hat X}_{\text{dS}};Q_\pm) {\tilde K}_{\Delta-\tfrac{d}2}\left(m\, R\right), \\
   &  X \in {\cal D}_-: \quad  \phi^\pm_{\Delta}(X;Q_\pm) =\mathcal{N}_\Delta G_{\Delta}(-e^{-\frac{\pi i}{2}}{\hat X}_{\text{dS}};Q_\pm) {\tilde K}_{\Delta-\tfrac{d}2}\left(m\, R\right).
\end{align}
\end{subequations}
 On a given hyperbolic slice of $\mathbb{M}^{d+2}$, the conformal primary wave functions $\phi^\pm_{\Delta}(X;Q_\pm)$ can be recast as bulk-to-boundary propagators with the same conformal weight $\Delta$ on $H^+_{d+1}$. In regions ${\cal A}_\pm$, using the anti-podal identification \eqref{apid} of $Q_+$ and $Q_-$, it is straightforward to see that: 
\begin{subequations}\label{phipmapm}
\begin{align}
& X \in {\cal A}_+: \quad   \phi^\pm_{\Delta}(X;Q_\pm) =  \mathcal{N}_\Delta G_{\Delta}({\hat X}_{\text{AdS}};Q_+) {\tilde K}_{\Delta-\frac{d}{2}}\left(m\, t e^{\pm \frac{\pi i}{2}}\right), \\
  & X \in {\cal A}_-: \quad   \phi^\pm_{\Delta}(X;Q_\pm) =  \mathcal{N}_\Delta G_{\Delta}({\hat X}_{\text{AdS}};Q_+) {\tilde K}_{\Delta-\tfrac{d}2}\left(m\, |t| e^{\mp \frac{\pi i}{2}}\right).
\end{align}
\end{subequations}
In region ${\cal D}_\pm$, the relation to the corresponding bulk-to-boundary propagator on $H^+_{d+1}$ is more subtle owing to the presence of a short distance singularity in $\phi^\pm_{\Delta}(X;Q_\pm)$ and an anti-podal singularity in $\phi^\mp_{\Delta}(X;Q_\mp)$. In global coordinates \eqref{dScoords}, we have
\begin{subequations}
\begin{align}
      \phi^+_{\Delta}(X;Q_+) &= e^{-\Delta \frac{\pi i}{2}}  \frac{\mathcal{N}_\Delta\, C^{\text{AdS}}_\Delta}{\left[2 \left(\text{sinh}\tau-\text{cosh}\tau \cos \gamma\right)-i \epsilon\right]^\Delta} {\tilde K}_{\Delta-\frac{d}{2}}\left(m R\right), \\
   \phi^-_{\Delta}(X;Q_-) &=  e^{-\Delta \frac{\pi i}{2}}  \frac{\mathcal{N}_\Delta\, C^{\text{AdS}}_\Delta}{\left[-2 \left(\text{sinh}\tau+\text{cosh}\tau \cos \gamma\right)-i \epsilon\right]^\Delta} {\tilde K}_{\Delta-\tfrac{d}2}\left(m R\right),
\end{align}
\end{subequations}
where 
\begin{equation}
    Q_\pm = \left(\pm 1,{\hat n}^\prime\right), \qquad \cos \gamma = {\hat n} \cdot {\hat n}^\prime.
\end{equation}
Let us first consider region ${\cal D}_+$, where $\phi^+_{\Delta}$ exhibits a short-distance singularity and $\phi^-_{\Delta}$ an anti-podal singularity. One zooms in on the short-distance singularity by taking the limit $\tau \to +\infty$ and $\gamma \to 0$:
\begin{subequations}
\begin{align}
    \phi^+_{\Delta}(X;Q_+) &= e^{-\Delta \frac{\pi i}{2}}  \frac{\mathcal{N}_\Delta\, C^{\text{AdS}}_\Delta}{\left[-2e^{-\tau}+\frac{1}{2}\gamma^2e^{\tau} -i \epsilon\right]^\Delta} {\tilde K}_{\Delta-\frac{d}{2}}\left(m R\right),\\
    &= e^{-\Delta \frac{\pi i}{2}} \mathcal{N}_\Delta\, C^{\text{AdS}}_\Delta \left(  \frac{\eta_+}{\eta^2_++|\vec{x}-\vec{x}^{\,\prime}|^2-i\epsilon}\right)^\Delta{\tilde K}_{\Delta-\frac{d}{2}}\left(m R\right),\label{phipsd}
\end{align}
\end{subequations}
where in the second equality we switched to Poincar\'e coordinates \eqref{poincoords} through the identifications $\eta_+ = 2e^{-\tau}$ and $|\vec{x}-\vec{x}^{\,\prime}| = \gamma$. For the anti-podal singularity of $\phi^-_{\Delta}$, one instead takes the limit $\tau \to +\infty$ and ${\bar \gamma} \to 0$ where ${\bar \gamma} = \gamma+\pi$:
\begin{subequations}
\begin{align}
    \phi^-_{\Delta}(X;Q_-) &= e^{-\Delta \frac{\pi i}{2}}  \frac{\mathcal{N}_\Delta\, C^{\text{AdS}}_\Delta}{\left[2e^{-\tau}-\frac{1}{2}\gamma^2e^{\tau} -i \epsilon\right]^\Delta} {\tilde K}_{\Delta-\frac{d}{2}}\left(m R\right),\\
    &= e^{-\Delta \frac{\pi i}{2}} \mathcal{N}_\Delta\, C^{\text{AdS}}_\Delta \left(  \frac{\eta_+}{\eta^2_+-|\vec{x}-\vec{x}^{\,\prime}|^2-i\epsilon}\right)^\Delta{\tilde K}_{\Delta-\frac{d}{2}}\left(m R\right).\label{phimap}
\end{align}
\end{subequations}
In the vicinity of the short-distance (s-d) and anti-podal (a-p) singularities in ${\cal D}_+$ we therefore have, respectively:\footnote{Note that $\phi^-_{\Delta}$ is analytic around the short distance singularity of $\phi^+_{\Delta}$ in ${\cal D}_+$. Likewise, $\phi^+_{\Delta}$ is analytic around the anti-podal singularity of $\phi^-_{\Delta}$. The analogous holds in region ${\cal D}_-$, which is considered below.} 
\begin{subequations}\label{phipmdp}
\begin{align}
   & \hspace*{-0.35cm}\text{s-d in ${\cal D}_+$:} \quad \phi^\pm_{\Delta}(X;Q_\pm) = e^{\Delta\left(1\mp 1\right)\frac{\pi i}{2}} {\cal N}_{\Delta}G_{\Delta}\left(z,\vec{x};\vec{x}^{\,\prime}\right){\tilde K}_{\Delta-\frac{d}{2}}\left(m R\right), \quad z = \eta_+ e^{- \frac{\pi i}{2}},\\
   & \hspace*{-0.35cm} \text{a-p in ${\cal D}_+$:} \quad \phi^\pm_{\Delta}(X;Q_\pm) = e^{-\Delta\left(1\pm 1\right)\frac{\pi i}{2}} {\cal N}_{\Delta}G_{\Delta}\left(z,\vec{x};\vec{x}^{\,\prime}\right){\tilde K}_{\Delta-\frac{d}{2}}\left(m R\right), \quad z = \eta_+ e^{+ \frac{\pi i}{2}},
\end{align}
\end{subequations}
where in Poincar\'e coordinates \eqref{poincoords}, the bulk-to-boundary propagator \eqref{buboEAdS} on $H^+_{d+1}$ reads:
\begin{equation}\label{bubopoinc}
    G_{\Delta}\left(z,\vec{x};\vec{x}^{\,\prime}\right) = C^{\text{AdS}}_\Delta\,\left(  \frac{z}{z^2+|\vec{x}-\vec{x}^{\,\prime}|^2}\right)^\Delta.
\end{equation}

\vskip 4pt
A similar result follows in region ${\cal D}_-$. Here, it is $\phi^-_{\Delta}$ that exhibits a short-distance singularity and $\phi^+_{\Delta}$ an anti-podal singularity. One zooms in on the short-distance singularity of $\phi^-_{\Delta}$ by taking the limit $\tau \to -\infty$ and $\gamma \to 0$:
\begin{subequations}
\begin{align}
    \phi^-_{\Delta}(X;Q_-) &= e^{-\Delta \frac{\pi i}{2}}  \frac{\mathcal{N}_\Delta\, C^{\text{AdS}}_\Delta}{\left[-2e^{\tau}+\frac{1}{2}\gamma^2e^{-\tau} -i \epsilon\right]^\Delta} {\tilde K}_{\Delta-\frac{d}{2}}\left(m R\right),\\
    &= e^{-\Delta \frac{\pi i}{2}} \mathcal{N}_\Delta\, C^{\text{AdS}}_\Delta \left(  \frac{\eta_-}{\eta^2_-+|\vec{x}-\vec{x}^{\,\prime}|^2-i\epsilon}\right)^\Delta{\tilde K}_{\Delta-\frac{d}{2}}\left(m R\right),\label{phipsd}
\end{align}
\end{subequations}
where in the second equality we introduced Poincar\'e coordinates \eqref{poincoords} through the identifications $\eta_- = 2e^{\tau}$ and $|\vec{x}-\vec{x}^{\,\prime}| = \gamma$. For the anti-podal singularity of $\phi^+_{\Delta}$ one instead takes the limit $\tau \to -\infty$ and ${\bar \gamma} \to 0$ where ${\bar \gamma} = \gamma+\pi$:
\begin{subequations}
\begin{align}
    \phi^+_{\Delta}(X;Q_+) &= e^{-\Delta \frac{\pi i}{2}}  \frac{\mathcal{N}_\Delta\, C^{\text{AdS}}_\Delta}{\left[2e^{+\tau}-\frac{1}{2}\gamma^2e^{-\tau} -i \epsilon\right]^\Delta} {\tilde K}_{\Delta-\frac{d}{2}}\left(m R\right),\\
    &= e^{-\Delta \frac{\pi i}{2}} \mathcal{N}_\Delta\, C^{\text{AdS}}_\Delta \left(  \frac{\eta_-}{\eta^2_--|\vec{x}-\vec{x}^{\,\prime}|^2-i\epsilon}\right)^\Delta{\tilde K}_{\Delta-\frac{d}{2}}\left(m R\right).\label{phimap}
\end{align}
\end{subequations}
By comparing with \eqref{bubopoinc}, we can conclude that in the vicinity of the short-distance (s-d) and anti-podal (a-p) singularities in ${\cal D}_-$ we have, respectively:
\begin{subequations}\label{phipmdm}
\begin{align}
   & \hspace*{-0.5cm}\text{s-d in ${\cal D}_-$:} \quad \phi^\pm_{\Delta}(X;Q_\pm) = e^{\Delta\left(1\pm 1\right)\frac{\pi i}{2}} {\cal N}_{\Delta}G_{\Delta}\left(z,\vec{x};\vec{x}^{\,\prime}\right){\tilde K}_{\Delta-\frac{d}{2}}\left(m R\right), \quad z = \eta_- e^{- \frac{\pi i}{2}},\\
   & \hspace*{-0.5cm} \text{a-p in ${\cal D}_-$:} \quad \phi^\pm_{\Delta}(X;Q_\pm) = e^{-\Delta\left(1\mp 1\right)\frac{\pi i}{2}} {\cal N}_{\Delta}G_{\Delta}\left(z,\vec{x};\vec{x}^{\,\prime}\right){\tilde K}_{\Delta-\frac{d}{2}}\left(m R\right), \quad z = \eta_- e^{+ \frac{\pi i}{2}}.
\end{align}
\end{subequations}
 
As we shall see in the following section, the relations \eqref{phipmapm}, \eqref{phipmdp} and \eqref{phipmdm} imply that contact diagram contributions to celestial correlators can be recast as contact Witten diagrams on $H^+_{d+1}$ upon integrating out the curvature radius of the hyperbolic slicing! This in turn leads to a more general relation between perturbative celestial correlators involving particle exchanges and corresponding Witten diagrams, which is presented in section \ref{subsec::allorders}.

\subsection{Contact diagrams} 
\label{subsec::contact}

\begin{figure}[htb]
    \centering
    \includegraphics[width=0.75\textwidth]{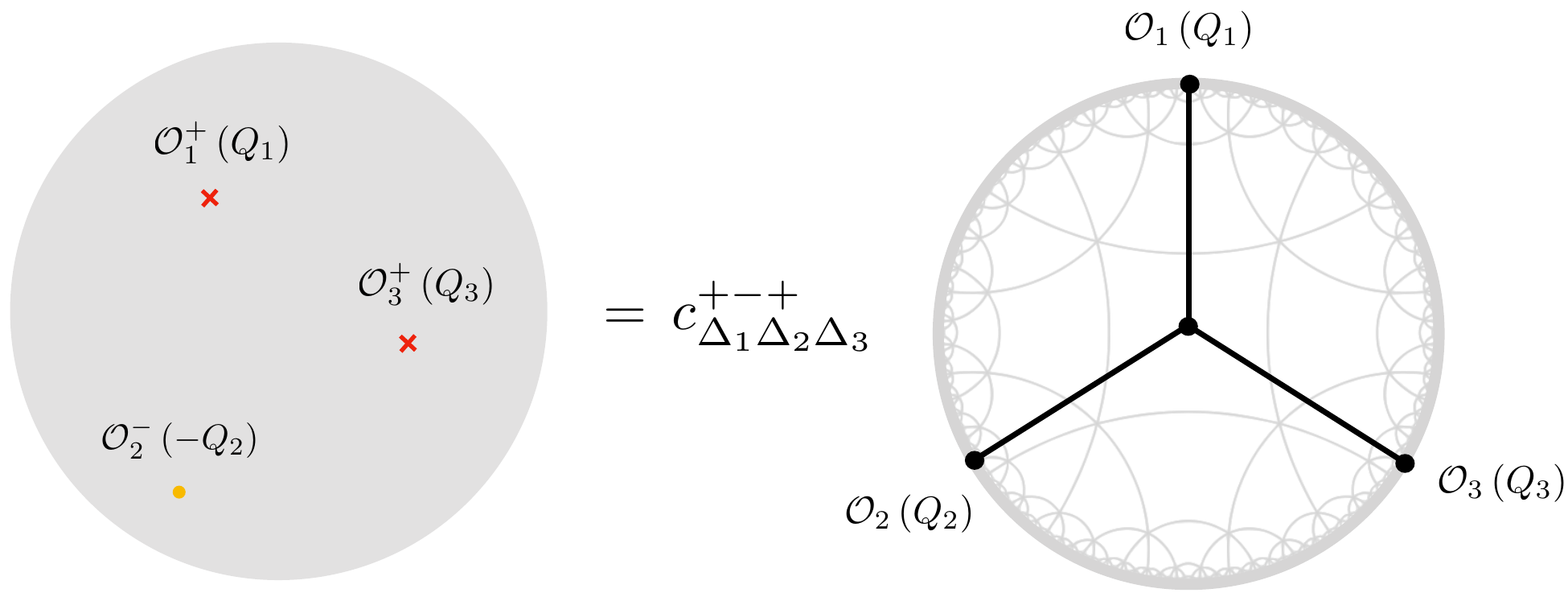}
    \caption{Contact diagram contributions to correlators on the $d$-dimensional celestial sphere (in grey) are proportional to their Witten diagram counterparts in EAdS$_{d+1}$ with proportionality constant \eqref{coeffnptfinal}. Red crosses denote operators inserted on the future boundary $S^+_d$ and yellow dots operators inserted on the past boundary boundary $S^-_d$. ${\cal O}^\pm_i$ is the conformal operator dual to the conformal primary wave function $\phi^\pm_i$.}
    \label{fig::contact3pt}
\end{figure}

Consider a theory of scalar fields $\phi_i$, $i=1,\ldots,n$ of mass $m_i$ interacting through the vertex
\begin{equation}
    {\cal V}\left(X\right) = g \phi_1\left(X\right)\ldots \phi_n\left(X\right).
\end{equation}
To leading order in $\lambda$ the $n$-point celestial correlator is given by
\begin{equation}\label{ccorrnpt}
 {\tilde {\cal A}}_{\Delta_1\ldots \Delta_n}^{\text{c}}\left(\pm_1 Q_1,\ldots,\pm_n Q_n\right) = -ig \int d^{d+2}X\, \phi^{\pm_1}_{\Delta_1}\left(X,\pm_1 Q_1\right)\ldots \phi^{\pm_n}_{\Delta_n}\left(X,\pm_n Q_n\right), \nonumber
\end{equation}
where the superscript ``c" to indicates that we are considering a contact diagram. The expressions \eqref{CPWFanals} for the conformal primary wave function instruct us to divide the integral over $\mathbb{M}^{d+2}$ into integrals over regions ${\cal A}_\pm$ and ${\cal D}_\pm$,
\begin{equation}
    \int d^{d+2}X = \int_{{\cal A}_+} d^{d+2}X+\int_{{\cal A}_-} d^{d+2}X+\int_{{\cal D}_+} d^{d+2}X+\int_{{\cal D}_-} d^{d+2}X,
\end{equation}
which in Poincar\'e coordinates \eqref{poincoords} read
\begin{subequations}
\begin{align}
    \int_{{\cal A}_+} d^{d+2}X &=  \int^\infty_0 t^{d+1} dt \int^\infty_0 \frac{dz}{z^{d+1}}\int d\vec{x},\\
    \int_{{\cal A}_-} d^{d+2}X &= \int^0_{-\infty} |t|^{d+1} dt \int^\infty_0 \frac{dz}{z^{d+1}}\int d\vec{x},\\
    \int_{{\cal D}_+} d^{d+2}X &= \int^\infty_0 R^{d+1} dR \int^\infty_0 \frac{d\eta_+}{\eta^{d+1}_+}\int d\vec{x},\\
    \int_{{\cal D}_-} d^{d+2}X &= \int^\infty_0 R^{d+1} dR \int^\infty_0 \frac{d\eta_-}{\eta^{d+1}_-}\int d\vec{x}.
\end{align}
\end{subequations}
Using the relations \eqref{phipmapm}, \eqref{phipmdp} and \eqref{phipmdm} between conformal primary wave functions and bulk-to-boundary propagators of the same conformal weight on $H^+_{d+1}$, the contribution from each region of $\mathbb{M}^{d+2}$ factorises into the corresponding $n$-point contact Witten diagram on $H^+_{d+1}$ and an integral over the curvature radius the hyperbolic foliation:\footnote{Here we replaced: 
\begin{equation}
    \int_{H^+_{d+1}} d{\hat X}_{\text{AdS}} = \int^\infty_0 \frac{dz}{z^{d+1}}\int d\vec{x}.
\end{equation}}
\begin{multline}\label{ccorrDfunc}
    {\tilde {\cal A}}_{\Delta_1\ldots \Delta_n}^{\text{c}}\left(\pm_1 Q_1,\ldots,\pm_n Q_n\right)= \underbrace{\left(c^{\pm_1 \ldots \pm_n}_{{\cal A}_+}+c^{\pm_1 \ldots \pm_n}_{{\cal A}_-}+c^{\pm_1 \ldots \pm_n}_{{\cal D}}\right)}_{c^{\pm_1 \ldots \pm_n}_{\Delta_1 \ldots \Delta_n}}\\  \times \underbrace{\int_{H^+_{d+1}} d{\hat X}_{\text{AdS}}\,G_{\Delta_1}({\hat X}_{\text{AdS}}; Q_1) \ldots G_{\Delta_n}({\hat X}_{\text{AdS}}; Q_n)}_{{}^{\left(\text{AdS}\right)}{\tilde {\cal A}}_{\Delta_1\ldots \Delta_n}^{\text{c}}\left( Q_1,\ldots,Q_n\right)}.
\end{multline}
The Witten diagram ${}^{\left(\text{AdS}\right)}{\tilde {\cal A}}_{\Delta_1\ldots \Delta_n}^{\text{c}}\left( Q_1,\ldots,Q_n\right)$ on the second line of \eqref{ccorrDfunc} is simply the well-known $D$-function \cite{DHoker:1999kzh}, which is defined as the $n$-point contact Witten diagram generated by a the non-derivative $n$-point interaction in $H^+_{d+1}$. The coefficient $c^{\pm_1 \ldots \pm_n}_{\Delta_1 \ldots \Delta_n}$ sums the radial integrals from each of the regions ${\cal A}_\pm$ and ${\cal D}$. For ${\cal A}_\pm$, using \eqref{phipmapm} we have:
\begin{subequations}\label{capam}
\begin{align}
    c^{\pm_1 \ldots \pm_n}_{{\cal A}_+}&=-ig \int^\infty_0 dt\, t^{d+1} \, \prod^n_{i=1}\,\mathcal{N}_{\Delta_i}\, {\tilde K}_{\Delta_i-\frac{d}{2}}\left(m_i\,t e^{\pm_i \frac{\pi i}{2}}\right),\\
    c^{\pm_1 \ldots \pm_n}_{{\cal A}_-}&=-ig\int^\infty_0 dt\, t^{d+1} \, \prod^n_{i=1}\,\mathcal{N}_{\Delta_i}\, {\tilde K}_{\Delta_i-\frac{d}{2}}\left(m_i\,t e^{\mp_i \frac{\pi i}{2}}\right).
\end{align} 
\end{subequations}
For region ${\cal D}$, let us first consider the case that the celestial correlator contains at least one incoming and at least one outgoing mode. In this case the integral over regions ${\cal D}_\pm$ receives contributions from both short-distance and anti-podal singularities. Inserting \eqref{phipmdp} and \eqref{phipmdm}, we have:
\begin{subequations}
\begin{align}
c^{\pm_1 \ldots \pm_n}_{{\cal D}}&=c^{\pm_1 \ldots \pm_n}_{{\cal D}_+}+c^{\pm_1 \ldots \pm_n}_{{\cal D}_-}\\
   c^{\pm_1 \ldots \pm_n}_{{\cal D}_+}
   &=-ig \left(\overbrace{\exp\left[\left(-d+\sum\limits^n_{j=1}\Delta_j\left(1\mp_j 1\right)\right)\frac{\pi i}{2}\right]}^{\text{s-d}\, \eqref{phipmdp}}+\overbrace{\exp\left[\left(d-\sum\limits^n_{j=1}\Delta_j\left(1\pm_j 1\right)\right)\frac{\pi i}{2}\right]}^{\text{a-p}\, \eqref{phipmdp}} \right) \nonumber \\ & \hspace*{5cm} \times \int^\infty_0 dR\, R^{d+1} \, \prod^n_{i=1}\,\mathcal{N}_{\Delta_i}\, {\tilde K}_{\Delta_i-\tfrac{d}2}\left(m_i\,R \right),\\
   c^{\pm_1 \ldots \pm_n}_{{\cal D}_-}&=-ig \left(\overbrace{\exp\left[\left(-d+\sum\limits^n_{j=1}\Delta_j\left(1\pm_j 1\right)\right)\frac{\pi i}{2}\right]}^{\text{s-d\,\eqref{phipmdm}}}+\overbrace{\exp\left[\left(d-\sum\limits^n_{j=1}\Delta_j\left(1\mp_j 1\right)\right)\frac{\pi i}{2}\right]}^{\text{a-p\,\eqref{phipmdm}}} \right) \nonumber \\ & \hspace*{5cm} \times \int^\infty_0 dR\, R^{d+1} \, \prod^n_{i=1}\,\mathcal{N}_{\Delta_i}\, {\tilde K}_{\Delta_i-\tfrac{d}2}\left(m_i\,R \right),
\end{align}
\end{subequations}
where the $d$-dependence in the phases comes from the change of integration variable from $\eta_\pm$ to $z$. These combine to give 
\begin{align}\label{CD12..n}
    c^{\pm_1 \ldots \pm_n}_{{\cal D}}
    &=-2ig\left(\overbrace{\cos\left[\left(d-\sum\limits^n_{j=1}\Delta_j\left(1\mp_j 1\right)\right)\frac{\pi}{2}\right]}^{\text{singularities of the $\phi^{+}_{\Delta_j}$}}+\overbrace{\cos\left[\left(d-\sum\limits^n_{j=1}\Delta_j\left(1\pm_j 1\right)\right)\frac{\pi}{2}\right]}^{\text{singularities of the $\phi^{-}_{\Delta_j}$}}\right),\nonumber \\ & \hspace*{5cm} \times \int^\infty_0 dR\, R^{d+1} \, \prod^n_{i=1}\,\mathcal{N}_{\Delta_i}\, {\tilde K}_{\Delta_i-\tfrac{d}2}\left(m_i\,R \right).
\end{align}

It is interesting to note that the integrals over the curvature radius $R$ and $t$ in $ c^{\pm_1 \ldots \pm_n}_{{\cal D}}$ and $c^{\pm_1 \ldots \pm_n}_{{\cal A}_\pm}$ are (analytic continuations) of the multiple-$K$ integrals often encountered in momentum space CFT \cite{Bzowski:2013sza,Bzowski:2020lip}. It is useful to study such integrals using the Mellin-Barnes representation \cite{Sleight:2019hfp,Sleight:2021iix,Sleight:2021plv}, where: 
\begin{subequations}\label{MBrepKtilde}
 \begin{align}
   \tilde{K}_{\Delta_j-\tfrac{d}{2}}(m_j R)&= \frac{1}{\Gamma\left(\Delta_j-\frac{d}{2}\right)}\int_{-i\infty}^{+i\infty}\frac{ds_j}{2\pi i}\ \tilde{K}_{\Delta_j-\tfrac{d}{2}}(m_j,s_j)R^{-\left(\frac{d}{2}+2 s_j\right)}\,,\\
   \tilde{K}_{\Delta_j-\tfrac{d}{2}}(m_j,s_j)&=\Gamma \left(s_j+\tfrac{1}{2}\left(\Delta_j-\tfrac{d}{2}\right)\right) \Gamma \left(s_j-\tfrac{1}{2}\left(\Delta_j-\tfrac{d}{2}\right)\right) \left(\frac{m}{2}\right)^{-2 s_j}.
\end{align}
 \end{subequations}
By adopting this representation the integral over the radial coordinate trivialises and is replaced by a Dirac delta function in the Mellin variables:
\begin{equation}\label{DDfunc}
    \int^\infty_0 dR\,R^{d+1} R^{-\sum\limits^n_{j=1}\left(2s_j+\frac{d}{2}\right)} = 2\pi i\, \delta\left(-\left(d+2\right)+\sum\limits^n_{j=1}\left(2s_j+\frac{d}{2}\right)\right).
\end{equation}
In the Mellin-Barnes representation, the coefficients \eqref{capam} differ simply by phases in the Mellin variables owing to the analytic continuation of the curvature radius:\footnote{Here we defined \begin{equation}
    \int^{+i\infty}_{-i\infty} \left[ds_j\right]_n = \int^{+i\infty}_{-i\infty} \frac{ds_1}{2\pi i}\ldots \frac{ds_n}{2\pi i}.
\end{equation}}
\begin{subequations}
\begin{align}
    c^{\pm_1 \ldots \pm_n}_{{\cal A}_+}&=-ig \, \int^{+i\infty}_{-i\infty} \left[ds_j\right]_n \, F_{\Delta_1 \ldots \Delta_n}\left(s_1,m_1,\ldots, s_n,m_n \right) e^{-\sum\limits^n_{j=1}\pm_j\left(\frac{d}{2}+2 s_j\right)\frac{\pi i}{2}},\\
    c^{\pm_1 \ldots \pm_n}_{{\cal A}_-}&=-ig\, \int^{+i\infty}_{-i\infty} \left[ds_j\right]_n \, F_{\Delta_1 \ldots \Delta_n}\left(s_1,m_1,\ldots, s_n,m_n \right) e^{\sum\limits^n_{j=1}\pm_j \left(\frac{d}{2}+2 s_j\right)\frac{\pi i}{2}},
\end{align}
\end{subequations}
where 
\begin{multline}\label{MBcontactAdS}
    F_{\Delta_1 \ldots \Delta_n}\left(s_1,m_1,\ldots, s_n,m_n \right)= 2\pi i\, \delta\left(-\left(d+2\right)+\sum\limits^n_{j=1}\left(2s_j+\tfrac{d}{2}\right)\right)\\ \times \prod^n_{i=1}\,\frac{\mathcal{N}_{\Delta_i}}{\Gamma\left(\Delta_i-\frac{d}{2}\right)}\, \tilde{K}_{\Delta_i-\tfrac{d}{2}}(m_i,s_i).
\end{multline}
The contributions from all four regions ${\cal A}_\pm$ and ${\cal D}_\pm$ then combine as:
\begin{align}\label{coeffnptfinal}
  c^{\pm_1 \ldots \pm_n}_{\Delta_1 \ldots \Delta_n} &= 2ig\, \int^{+i\infty}_{-i\infty} \left[ds_j\right]_n \, \mathfrak{c}^{\pm_1 \ldots \pm_n}\left(s_1,\ldots,s_n\right)\,  F_{\Delta_1 \ldots \Delta_n}\left(s_1,m_1,\ldots, s_n,m_n \right),
\end{align}
where
\begin{equation}\label{sincfrak}
    \mathfrak{c}^{\pm_1 \ldots \pm_n}\left(s_1,\ldots,s_n\right)\, =\mathfrak{c}^{\pm_1 \ldots \pm_n}_{{\cal A}_+}+\mathfrak{c}^{\pm_1 \ldots \pm_n}_{{\cal A}_-}+\mathfrak{c}^{\pm_1 \ldots \pm_n}_{{\cal D}},
\end{equation}
with\footnote{Here we simplified the cosine factors in \eqref{CD12..n} using the identity:
\begin{equation}
    2\cos \alpha\,\cos \beta = \cos\left(\alpha+\beta\right)+\cos\left(\alpha-\beta\right).
\end{equation}}
\begin{subequations}\label{ccpm}
\begin{align}\label{ccalapm}
    \mathfrak{c}^{\pm_1 \ldots \pm_n}_{{\cal A}_+}+\mathfrak{c}^{\pm_1 \ldots \pm_n}_{{\cal A}_-}&=
    -\cos\left[\sum\limits^n_{j=1}\pm_j \left(\frac{d}{2}+2 s_j\right)\frac{\pi }{2}\right],\\\mathfrak{c}^{\pm_1 \ldots \pm_n}_{{\cal D}}&=-2\cos\left[\left(d-\sum\limits^n_{j=1}\Delta_j\right)\frac{\pi}{2}\right]\cos\left[\left(\sum\limits^n_{j=1}\pm_j \Delta_j\right) \frac{\pi}{2}\right].\label{ccDpm}
\end{align}
\end{subequations}
The sinusoidal factor in the Mellin variables $s_j$, coming from regions ${\cal A}_\pm$, indicates the presence of folded singularities in the mass $m_j$ -- see \cite{Sleight:2021plv} section 3.1. Such folded singularities are absent in region ${\cal D}$, in which case the corresponding coefficient \eqref{ccDpm} is constant and is, in particular, given by a product of cosine factors. This in turn implies that the contribution from region ${\cal D}$ vanishes for certain values of $\Delta_j$, corresponding to zeros of the cosine factors in \eqref{ccDpm}. From the expressions \eqref{ccpm} we can infer the following useful properties of the coefficient $ c^{\pm_1 \ldots \pm_n}_{\Delta_1 \ldots \Delta_n}$:
\begin{subequations}
\begin{align}
    c^{\pm_1 \ldots \pm_n}_{\Delta_1 \ldots \Delta_n} &= c^{\mp_1 \ldots \mp_n}_{\Delta_1 \ldots \Delta_n},\\
    c^{\pm \ldots \mp\ldots \pm}_{\Delta_1 \ldots \Delta_j\ldots \Delta_n} &= -c^{\pm \ldots \mp\ldots \pm}_{\Delta_1 \ldots d-\Delta_j\ldots \Delta_n}.\label{symmcoeffc}
\end{align}
\end{subequations}
The second property states that correlators involving a single incoming or outgoing mode are anti-symmetric upon replacing the mode with its shadow: $\Delta_j \to d-\Delta_j$.\footnote{To see this one also uses that:
\begin{equation}
    \frac{{\cal N}_{\Delta}}{\Gamma\left(\Delta-\frac{d}{2}\right)} =-\frac{{\cal N}_{d-\Delta}}{\Gamma\left(\frac{d}{2}-\Delta\right)}.
\end{equation}} 

\vskip 4pt
It is interesting to compare the coefficient $ c^{\pm_1 \ldots \pm_n}_{\Delta_1 \ldots \Delta_n}$ (which, we recall, originates from integrating out the curvature radius in the hyperbolic slicing of $\mathbb{M}^{d+2}$) with momentum space $n$-point scalar contact Witten diagrams in EAdS. In particular, upon multiplying by $R^d$ and $m^{\Delta}$, the function $\tilde{K}_{\Delta-\tfrac{d}{2}}(m_j R)$ defined in \eqref{tildeK} is proportional to the bulk-to-boundary propagator in Poincar\'e coordinates of $H^+_{d+1}$. Here the mass $m$ plays the role of the magnitude of the boundary momentum and $R$ the role of the Poincar\'e coordinate $z$. The difference is that, in regions ${\cal A}_+$ and ${\cal A}_-$, this coordinate is Wick rotated according to whether a particle is incoming or outgoing, leading to folded singularities in the ``momenta" $m_i$. Such folded singularities are absent in standard momentum space EAdS Witten diagrams. In the Mellin-Barnes representation, these singularities manifest themselves in the sinusoidal factor \eqref{ccalapm} in the Mellin variables, where \eqref{MBcontactAdS} is, up to factors of $m^{\Delta_j}$, indeed proportional to the Mellin-Barnes representation of an $n$-point contact Witten diagram in EAdS \cite{Sleight:2021plv}. It is interesting to note that the folded singularities only originate from the regions ${\cal A}_\pm$, where $\mathbb{M}^{d+2}$ is foliated by EAdS$_{d+1}$ hypersurfaces. The contribution \eqref{ccDpm} from region ${\cal D}$, where $\mathbb{M}^{d+2}$ is instead foliated by dS$_{d+1}$ hypersurfaces, does not contain folded singularities and, up to factors of $m^{\Delta_j}$, takes the same form as momentum space $n$-point contact Witten diagrams in EAdS.

\vskip 4pt 
Let us now turn to the case that we have all incoming or all outgoing modes. In this case there is just a single contribution from ${\cal D}_+$, which, in the case that we have all outgoing modes, corresponds to the short distance singularity of $\phi^+_{\Delta}$  or, in the case that we have all incoming modes, the anti-podal singularity of $\phi^-_{\Delta}$. And likewise for ${\cal D}_-$ which, in the case that we have all outgoing modes, corresponds to the anti-podal singularity of $\phi^+_{\Delta}$ or, in the case that we have all incoming modes, the short-distance singularity of $\phi^-_{\Delta}$. There is therefore only a single cosine factor in \eqref{CD12..n} that contributes in this case, giving:
\begin{equation}\label{contactvanish}
    \mathfrak{c}^{\pm \ldots \pm}\left(s_1,\ldots,s_n\right)\, =-\left[\cos\left( \frac{d\pi }{2}\right)+\cos\left( \left(d+2\right)\frac{\pi }{2}\right)\right]=0,
\end{equation}
where we used the Dirac delta function \eqref{DDfunc} to simplify the total contribution \eqref{ccalapm} from ${\cal A}_+$ and ${\cal A}_-$. This vanishing is consistent with the fact that there is no scattering amplitude with all incoming or all outgoing particles. This in particular means that there is no non-trivial celestial two-point function with two incoming or two outgoing particles.

 \vskip 4pt
 It should be noted that, while for generic values of the scaling dimensions $\Delta_i$ the need for a Mellin-Barnes representation \eqref{coeffnptfinal} of the coefficient $c^{\pm_1 \ldots \pm_n}_{\Delta_1 \ldots \Delta_n}$ is inevitable, for certain values of $\Delta_i$ the Mellin-Barnes integrals can be lifted. See e.g. sections 3.3-3.4 in \cite{Sleight:2019mgd} and section 4.6 of \cite{Sleight:2019hfp}. An example is when $\Delta_i=\frac{d+1}{2}$ which corresponds to a conformally coupled scalar. In the following section we study this case in more detail for $n=3$.
 
\paragraph{Specific cases: 3pt contact diagrams.}

In the above we saw that celestial $n$-point contact diagrams can be re-expressed \eqref{ccorrDfunc} as an $n$-point contact Witten diagram in $H^+_{d+1}$ times a coefficient $c^{\pm_1 \ldots \pm_n}_{\Delta_1 \ldots \Delta_n}$ that arises from the integral over the curvature radius of the hyperbolic slicing. In the following we analyse in more detail the case $n=3$, where we have that 
\begin{subequations}
\begin{align}
    \mathfrak{c}^{+++}(s_i)&=\mathfrak{c}^{---}(s_i)=0\,,\\ \label{c1}
    \mathfrak{c}^{-++}(s_i)&=\mathfrak{c}^{+--}(s_i)=\cos(2\pi s_1)-\cos\left[\left(\tfrac{d}{2}-\Delta_1\right) \pi\right]-\cos\left[\left(\tfrac{d}{2}-\Delta_2-\Delta_3\right) \pi\right]\,,\\ \label{c2}
    \mathfrak{c}^{+-+}(s_i)&=\mathfrak{c}^{-+-}(s_i)=\cos(2\pi s_2)-\cos\left[\left(\tfrac{d}{2}-\Delta_2\right) \pi\right]-\cos\left[\left(\tfrac{d}{2}-\Delta_1-\Delta_3\right) \pi\right]\,,\\ \label{c3}
    \mathfrak{c}^{++-}(s_i)&=\mathfrak{c}^{--+}(s_i)=\cos(2\pi s_3)-\cos\left[\left(\tfrac{d}{2}-\Delta_3\right) \pi\right]-\cos\left[\left(\tfrac{d}{2}-\Delta_1-\Delta_2\right) \pi\right]\,.
\end{align}
\end{subequations}
For generic $\Delta_i$, the Mellin-Barnes integral \eqref{coeffnptfinal} with $n=3$ can be expressed in terms of the Appell $F_4$ function (see appendix B.2 of \cite{Sleight:2019mgd}). For certain values of the scaling dimensions, however, the corresponding the Mellin-Barnes integrals can be lifted. The simplest example is for conformally coupled scalars, corresponding to $\Delta_i=\frac{d+1}2$ (see section 3.3 of \cite{Sleight:2019mgd}), which can be used as seeds for correlators of conserved currents (as in e.g. \cite{Sleight:2019hfp,Sleight:2021iix}). Considering for concreteness the case that we have one incoming mode and two outgoing (i.e. $\pm_1 = +$, $\pm_2 = +$ and $\pm_3 = -$), upon evaluating all the Mellin-Barnes integrals one obtains:
\begin{subequations}
\begin{align}
     c^{++-}_{\frac{d+1}2 \frac{d+1}2 \frac{d+1}2|\mathcal{A}} &= i g \frac{ \cos\left(\frac{d \pi}{2}\right) \Gamma \left(\frac{1-d}{2}\right)  }{\sqrt{2m_1 m_2 m_3}}\left(m_3-m_1-m_2\right)^{\frac{d-1}{2}},\\
     c^{++-}_{\frac{d+1}2 \frac{d+1}2 \frac{d+1}2|\mathcal{D}}&=i g \frac{ \cos\left(\frac{d \pi}{2}\right) \Gamma \left(\frac{1-d}{2}\right)  }{\sqrt{2m_1 m_2 m_3}}\left(m_1+m_2+m_3\right)^{\frac{d-1}{2}},
\end{align}
\end{subequations}
where for convenience we gave the contributions from region ${\cal D}$ and combined regions ${\cal A}_\pm$ separately. Note the expected folded singularities in the mass $m_i$ coming from regions ${\cal A}_\pm$. Setting $m_1=m_2=m$, $m_3=2m(1+\epsilon)$ and $d=2$, the contribution from regions ${\cal A}_\pm$ recovers the result given in equation (3.13) of \cite{Pasterski:2016qvg} which was obtained from the corresponding momentum space scattering amplitude using \eqref{ccorr}. We therefore differ from the result of \cite{Pasterski:2016qvg} by the contribution from region ${\cal D}$, which is regular in $\epsilon$.\footnote{It should be clarified from where this discrepancy originates, which we leave to future work. A possible explanation could be that employing \eqref{ccorr} to obtain celestial correlators from momentum space scattering amplitudes requires to commute the divergent momentum integral past the integral over the bulk of Minkowski space.}

\vskip 4pt
Similarly, one can consider the case that one has two conformally coupled and one general scalar.  The fact we have two conformally coupled scalars implies that two of the three Mellin-Barnes integrals can be lifted. One obtains the following expression in terms of the Gauss hypergeometric function:
\begin{subequations}
\begin{align}
    c^{++-}_{\frac{d+1}2 \frac{d+1}2 \Delta_3|\mathcal{A}}&=-i\cos \left(\tfrac{d \pi  }{2}\right)\frac{\sqrt{\pi } g \left(2m_3\right)^{\frac{d-1}{2}}  }{\sqrt{2m_1m_2m_3}\, \Gamma (\Delta_3 -\frac{d}{2})\Gamma\left(\frac{3-d}{2}\right)}\, \\ \nonumber
    & \hspace*{4cm} \times _2F_1\left(1-\Delta_3,1-d+\Delta_3;\frac{3-d}{2};\frac{m_1+m_2+m_3}{2 m_3}\right),\\
    c^{++-}_{\frac{d+1}2 \frac{d+1}2 \Delta_3|\mathcal{D}}&=-i\left(\cos \left(\tfrac{d \pi  }{2}\right)-\cos \left(\left(\tfrac{d}{2}-\Delta_3\right)\pi\right)\right)\frac{\sqrt{\pi } g \left(2m_3\right)^{\frac{d-1}{2}}}{\sqrt{2m_1m_2m_3}\, \Gamma (\Delta_3 -\frac{d}{2})\Gamma\left(\frac{3-d}{2}\right)} \\ \nonumber
    & \hspace*{4cm} \times _2F_1\left(1-\Delta_3,1-d+\Delta_3;\frac{3-d}{2};\frac{m_3-m_1-m_2}{2 m_3}\right).
\end{align}
\end{subequations}
In $d=2$ this gives:
\begin{subequations}
\begin{align}
    c^{++-}_{\tfrac{3}{2}\tfrac{3}{2},1+i \nu|\mathcal{A}}&=2  ig\Gamma\left(1-\Delta_3\right)\frac{  \cos \left(2 \left(1-\Delta_3\right)  \csc ^{-1}\left(\frac{\sqrt{2m_3}}{\sqrt{m_1+m_2+m_3}}\right)\right)}{\sqrt{m_1m_2}}\,,\\
    c^{++-}_{\tfrac{3}{2}\tfrac{3}{2},1+i \nu|\mathcal{D}}&=2ig \left(1+\cos \left(\left(1-\Delta_3\right)\pi\right)\right)\Gamma\left(1-\Delta_3\right)\frac{  \text{cosh} \left(2 \left(1-\Delta_3\right)  \text{csch}^{-1}\left(\frac{ \sqrt{2m_3}}{\sqrt{m_1+m_2-m_3}}\right)\right)}{\sqrt{m_1m_2}}.
\end{align}
\end{subequations}

\vskip 4pt
Another example in which simplifications arise is when one or more of the fields are massless. In this case the function \eqref{tildeK} in the representation \eqref{cpwconcf} of the conformal primary wavefunction simplifies \cite{Pasterski:2017kqt}. In the case of two massless external fields we have: 
\begin{align}
    c^{++-}_{\tfrac{d}{2}+i\nu_1,\tfrac{d}{2}+i\nu_2,\tfrac{d}{2}+i\nu_3|\mathcal{A}}&=\frac{i g \cos \left(\pi  \left(\Delta_1+\Delta_2-\frac{d}{2}\right)\right) m_3^{\Delta_1+\Delta_2-d} \left(2m_3\right)^{2-\tfrac{d}{2}}}{\Gamma \left(\Delta_3-\frac{d}{2}\right)}\\ \nonumber
    & \hspace*{4cm}\times  \Gamma \left(1+\tfrac{\Delta_3-\Delta_1-\Delta_2}{2}\right) \Gamma \left(1+\tfrac{d-\Delta_1-\Delta_2-\Delta_3}{2}\right),\\ \nonumber
     c^{++-}_{\tfrac{d}{2}+i\nu_1,\tfrac{d}{2}+i\nu_2,\tfrac{d}{2}+i\nu_3|\mathcal{D}}&=\frac{i g \left(\cos\left[\left(\tfrac{d}{2}-\Delta_3\right) \pi\right]+\cos\left[\left(\Delta_1+\Delta_2-\tfrac{d}{2}\right) \pi\right]\right) m_3^{\Delta_1+\Delta_2-d} \left(2m_3\right)^{2-\tfrac{d}{2}}}{\Gamma \left(\Delta_3-\frac{d}{2}\right)}\\ 
    & \hspace*{3cm}\times  \Gamma \left(1+\tfrac{\Delta_3-\Delta_1-\Delta_2}{2}\right) \Gamma \left(1+\tfrac{d-\Delta_1-\Delta_2-\Delta_3}{2}\right).
\end{align}

\vskip 4pt
Let us finally note that these examples make clear that the coefficients \eqref{coeffnptfinal} are vanishing for certain mass configurations. Analogous zeros appear in the context of dS boundary correlators \cite{Sleight:2021plv} and, as in that case, we expect them to be imprints of (perturbative) unitarity.

\subsection{All orders in the perturbative expansion}
\label{subsec::allorders}

In the previous section we saw that contact diagram contributions to celestial correlators are proportional to the corresponding contact Witten diagram in $H^+_{d+1}$:
\begin{equation}\label{AdStocelestcont}
   {\tilde {\cal A}}^\text{c}_{\Delta_1\ldots \Delta_n}\left(\pm_1 Q_1,\ldots,\pm_n Q_n\right) =  c^{\pm_1 \ldots \pm_n}_{\Delta_1 \ldots \Delta_n}\, {}^{\left(\text{AdS}\right)}{\tilde {\cal A}}^\text{c}_{\Delta_1\ldots \Delta_n}\left(Q_1,\ldots, Q_n\right).
\end{equation} 
The coefficients $c^{\pm_1 \ldots \pm_n}_{\Delta_1 \ldots \Delta_n}$ can be thought of as the ratio:  
\begin{equation}\label{cratiolambda}
    c^{\pm_1 \ldots \pm_n}_{\Delta_1 \ldots \Delta_n} = \frac{\lambda^{\pm_1 \ldots \pm_n}_{\Delta_1 \ldots \Delta_n}}{\lambda^{\text{AdS}}_{\Delta_1 \ldots \Delta_n}},
\end{equation}
relating the coefficient $\lambda^{\text{AdS}}_{\Delta_1 \ldots \Delta_n}$ of contact Witten diagram to its celestial counterpart $\lambda^{\pm_1 \ldots \pm_n}_{\Delta_1 \ldots \Delta_n}$. 

\vskip 4pt
Contact diagrams are the basic building blocks for all other processes i.e. those that involve particle exchanges. When an exchanged particle goes on-shell, we expect that the corresponding observable factorises appropriately into subprocesses and, when all exchanged particles are on-shell, these subprocesses are the contact diagrams generated by the vertices. Using this property we can extend the result \eqref{ccorrDfunc} to celestial correlator involving exchanged particles, decomposing them as a sum of corresponding Witten diagrams in $H^+_{d+1}$.

\begin{figure}[htb]
    \centering
    \includegraphics[width=0.85\textwidth]{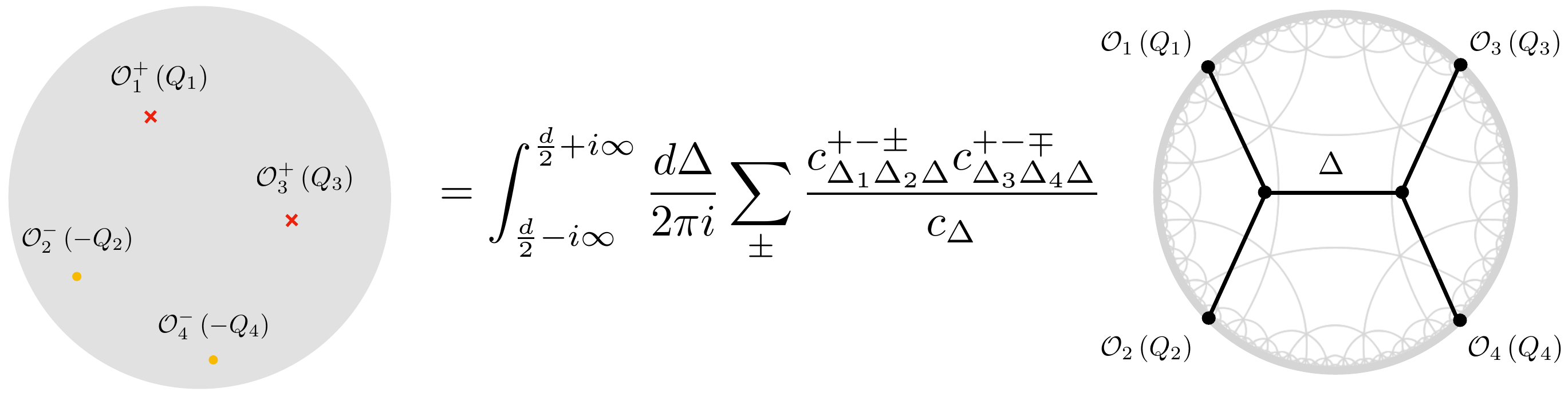}
    \caption{Contributions from particle exchanges to $d$-dimensional celestial correlators decompose into exchange Witten diagrams of particles carrying Principal Series representations $\Delta \in \frac{d}{2}+i\mathbb{R}$ in EAdS$_{d+1}$. The relative coefficients ensure consistent on-shell factorisation.}
    \label{fig::exch}
\end{figure}

\vskip 4pt
To illustrate, let us first consider the tree level exchange of a scalar field of mass $m^2$ in a 4pt celestial correlator, say in the $s$-channel. Using that a field in $\mathbb{M}^{d+2}$ can be decomposed \eqref{flatAdSdecomp} in terms of fields on (EA)dS$_{d+1}$ carrying principal series representations of $SO\left(d+1,1\right)$, assuming consistent on-shell factorisation then implies
\begin{multline} \label{cspexch}
    {\tilde {\cal A}}^{\text{exch.}}_{\Delta_1\Delta_2|m^2|\Delta_3\Delta_4}\left(\pm_1 Q_1,\pm_2 Q_2,\pm_3 Q_3,\pm_4 Q_4\right)\\=\int^{\frac{d}{2}+i\infty}_{\frac{d}{2}-i\infty} \frac{d\Delta}{2\pi i}\, \sum\limits_{\pm}\frac{c^{\pm_1 \pm_2 \pm}_{\Delta_1 \Delta_2 \Delta}c^{\mp \pm_3 \pm_4}_{\Delta \Delta_3 \Delta_4}}{c_{\Delta}}\,{}^{\left(\text{AdS}\right)}{\tilde {\cal A}}^{\text{exch.}}_{\Delta_1\Delta_2|\Delta|\Delta_3\Delta_4}\left(Q_1,Q_2,Q_3,Q_4\right),
\end{multline}
where ${}^{\left(\text{AdS}\right)}{\tilde {\cal A}}_{\Delta_1\Delta_2|\Delta|\Delta_3\Delta_4}\left(Q_1,Q_2,Q_3,Q_4\right)$ is the corresponding exchange Witten diagram in $H^+_{d+1}$ for a particle of scaling dimension $\Delta$. The coefficients $c^{\pm_1 \pm_2 \pm}_{\Delta_1 \Delta_2 \Delta}$ and $c^{\mp \pm_3 \pm_4}_{\Delta \Delta_3 \Delta_4}$ convert the coefficients of the three-point contact subdiagrams in AdS on the r.h.s. to their celestial counterparts via \eqref{cratiolambda}, where for each $\Delta$ there are two operators corresponding to incoming ($-$) and outgoing ($+$) particles.\footnote{Note that there is no two-point function for two incoming or two outgoing particles, which can be seen from \eqref{contactvanish}.} The $c_{\Delta}$ account for the change in 2pt function normalisation, 
\begin{align}\label{cseadstods}
   c_{\Delta} &=\frac{C^{\text{flat}}_{\Delta}}{C^{\text{AdS}}_{\Delta}}= \frac{\Gamma\left(\frac{d}{2}-\Delta\right)\Gamma\left(\Delta-\frac{d}{2}+1\right)}{2\pi}=\frac{1}{2}\csc\left(\left(\tfrac{d}{2}-\Delta\right)\pi\right),
   \end{align}
where $C^{\text{flat}}_{\Delta}$ is the celestial 2pt function normalisation corresponding to a canonical Klein-Gordon inner product \eqref{cKG} and the AdS normalisation $C^{\text{AdS}}_{\Delta}$ is given in \eqref{buboEAdS}. These clearly have the following useful property upon replacing $\Delta$ with its shadow $d-\Delta$:
\begin{equation}
    c_{\Delta} = - c_{d-\Delta}, \label{2ptshadcoeff}
\end{equation}
which is a specific case of the more general property \eqref{symmcoeffc}.

\vskip 4pt
More generally, given a perturbative contribution to an $n$-point celestial correlator \eqref{ccorr}, the following steps give its decomposition in terms of corresponding Witten diagrams in $H^+_{d+1}$:

\begin{enumerate}
    \item Draw the same diagram in EAdS: Each external line connected to the point $\pm_i Q_i$ on the conformal boundary of $\mathbb{M}^{d+2}$ becomes an external line connected to the point $Q_i$ on the boundary of EAdS$_{d+1}$, where an operator with the same scaling dimension $\Delta_i$ is inserted. Each internal line becomes an internal line in EAdS. Each vertex becomes a vertex in EAdS. 
    
    \item Assign each internal line a scaling dimension label and divide by the factor \eqref{cseadstods} accounting for the change in 2pt function normalisation. For each vertex, multiply by the factor \eqref{cratiolambda} that converts the contact diagram it generates to the corresponding celestial contact diagram as in \eqref{AdStocelestcont}. For vertices that are attached to internal lines, in the corresponding contact diagram the external particle can be either incoming or outgoing and these possibilities should be summed over.
    
    \item For each internal line, integrate the associated scaling dimension label over the Principal Series $\Delta \in \frac{d}{2}+i\mathbb{R}$.
\end{enumerate}

For example, for the candy diagram contribution to the four-point function in the $s$-channel, taking $\pm_1=\pm_3=+1$ and $\pm_2=\pm_4=-1$, one obtains:

\begin{equation}\nonumber
    \includegraphics[width=0.85\textwidth]{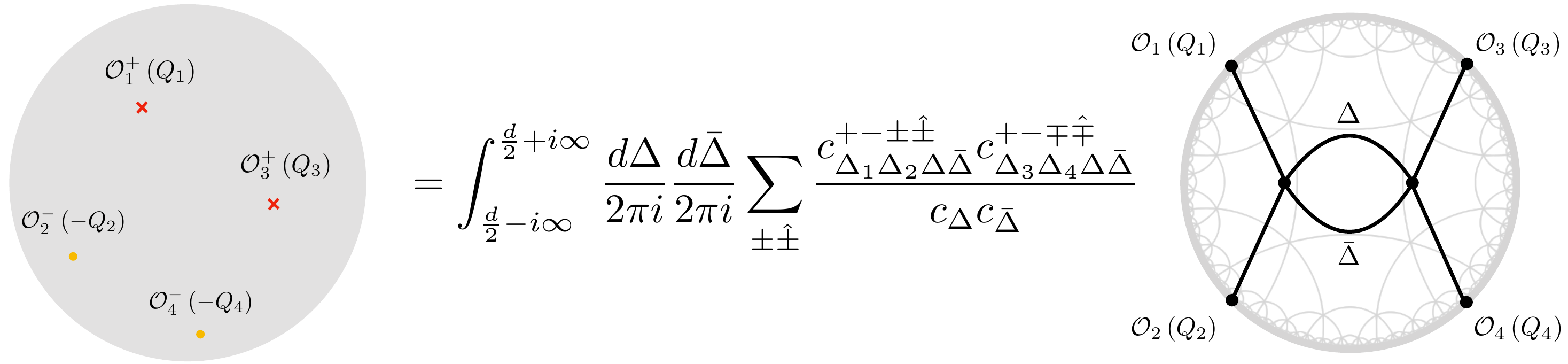}.
\end{equation}

\vskip 4pt
To summarise this section, we have seen that a given perturbative contribution to a celestial correlator can be expanded in terms of corresponding Witten diagrams in $H^+_{d+1}$. This was shown to follow from the assumption of consistent on-shell factorisation once it was established in section \ref{subsec::contact} that celestial contact diagrams are proportional their Witten diagram counterparts in $H^+_{d+1}$ (equation \ref{ccorrDfunc}). It should be noted that this result would equivalently follow from conformal symmetry, factorisation and the assumption of single-valuedness (analyticity) of celestial correlators as a function of the (complexified) cross-ratios, as is the case for AdS boundary correlators in the Euclidean region. The fact that celestial contact diagrams are proportional to their corresponding contact Witten diagrams in $H^+_{d+1}$ indeed shows that the assumption of single-valuedness would have been a valid one.

\subsection{Spectral representation and unitarity}
\label{subsec::spectralrep}

We have seen that perturbative celestial correlators in $d$-dimensions can be decomposed in terms of corresponding Witten diagrams on EAdS$_{d+1}$. Through such identities one might try to import the wealth of techniques, results and understanding available in EAdS to the study of celestial correlators, in the same spirit as analogous studies relating perturbative boundary correlators in dS$_{d+1}$ to their EAdS$_{d+1}$ counterparts \cite{Sleight:2020obc,Sleight:2021plv}. 

\vskip 4pt
For example, such identities imply that perturbative celestial correlators have the same analytic structure as their EAdS Witten diagram counterparts. In particular, given a single-valued conformally invariant four-point function of operators ${\cal O}_i$ in $d$-dimensional Euclidean space, harmonic analysis on $SO\left(d+1,1\right)$~\cite{Dobrev:1975ru,Dobrev:1977qv,Mack:2009mi,Costa:2012cb,Caron-Huot:2017vep} implies that it admits a partial-wave expansion over an orthogonal basis of Eigenfunctions ${\cal F}_{\Delta,J}$ of the Casimir invariants taking the following form (say, in the (12)(34) channel): 
\begin{equation}\label{cpwe}
   \hspace*{-0.4cm} \langle {\cal O}_1(\vec{x}_1){\cal O}_2(\vec{x}_2){\cal O}_3(\vec{x}_3){\cal O}_4(\vec{x}_4) \rangle={\mathds 1}_{12} {\mathds 1}_{34}+\sum\limits^\infty_{J=0}\int^{\frac{d}{2}+i\infty}_{\frac{d}{2}-i\infty} \frac{{\rm d}\Delta}{2\pi i}\,\rho_{J}(\Delta)\,{\cal F}^{12,34}_{\Delta,J}(\vec{x}_1,\vec{x}_2,\vec{x}_3,\vec{x}_4)\,,
\end{equation}
where $\rho_{J}(\Delta)$ is the spectral density and the first term on the r.h.s. is the contribution from the identity operator. For correlators \eqref{cpwe} on the boundary of EAdS $\rho_{J}(\Delta)$ is a meromorphic function of $\Delta$, which is due to the fact that the operator product expansion converges. In perturbation theory, celestial correlators can be expressed as a sum of EAdS Witten diagrams, implying that $\rho_{J}(\Delta)$ is meromorphic also for perturbative celestial correlators.

\vskip 4pt
The conformal partial wave expansion (CPWE) of celestial correlators has been studied in a variety of works \cite{Lam:2017ofc,Nandan:2019jas,Atanasov:2021cje,Melton:2021kkz}. The above results tell us that one can determine the CPWE of a given perturbative celestial correlator can be determined using their relation to EAdS Witten diagrams. For example, considering the exchange \eqref{cspexch} with $\pm_1=\pm_2=+1$ and $\pm_3=\pm_4=-1$, we can write:
\begin{align}  \nonumber
    {\tilde {\cal A}}^{\text{exch.}}_{\Delta_1\Delta_2|m^2|\Delta_3\Delta_4}&=\int^{\frac{d}{2}+i\infty}_{\frac{d}{2}-i\infty} \frac{d\Delta}{2\pi i}\,\frac{c^{+ + -}_{\Delta_1 \Delta_2 \Delta}c^{+ - -}_{\Delta \Delta_3 \Delta_4}}{c_{\Delta}}\,{}^{\left(\text{AdS}\right)}{\tilde {\cal A}}^{\text{exch.}}_{\Delta_1\Delta_2|\Delta|\Delta_3\Delta_4},\\
    &=\frac{1}{2}\int^{\frac{d}{2}+i\infty}_{\frac{d}{2}-i\infty} \frac{d\Delta}{2\pi i}\, \frac{c^{+ + -}_{\Delta_1 \Delta_2 \Delta}c^{+ - -}_{\Delta \Delta_3 \Delta_4}}{c_{\Delta}}\,\left[{}^{\left(\text{AdS}\right)}{\tilde {\cal A}}^{\text{exch.}}_{\Delta_1\Delta_2|\Delta|\Delta_3\Delta_4}-{}^{\left(\text{AdS}\right)}{\tilde {\cal A}}^{\text{exch.}}_{\Delta_1\Delta_2|d-\Delta|\Delta_3\Delta_4}\right], \nonumber
\end{align}
where in the first equality we used that $c^{+ + +}_{\Delta_1 \Delta_2 \Delta}=c^{- - -}_{\Delta \Delta_3 \Delta_4}=0$ and in second equality we used the anti-symmetric property \eqref{symmcoeffc} of the coefficients upon replacing $\Delta$ with its shadow $d-\Delta$. The difference of $\Delta$ and $d-\Delta$ exchanges in EAdS is proportional to a conformal partial wave \cite{Leonhardt:2003qu,Costa:2014kfa}:
\begin{equation}
    {}^{\left(\text{AdS}\right)}{\tilde {\cal A}}^{\text{exch.}}_{\Delta_1\Delta_2|\Delta,J|\Delta_3\Delta_4}-{}^{\left(\text{AdS}\right)}{\tilde {\cal A}}^{\text{exch.}}_{\Delta_1\Delta_2|d-\Delta,J|\Delta_3\Delta_4}=\left(d-2\Delta\right) {}^{\left(\text{AdS}\right)}{\cal F}^{12,34}_{\Delta,J}, 
\end{equation}
where, adopting the notation of \cite{Sleight:2021plv}, the AdS-normalised conformal partial wave reads:
\begin{equation}
    {}^{\left(\text{AdS}\right)}{\cal F}^{12,34}_{\Delta,J}(\vec{x}_1,\vec{x}_2,\vec{x}_3,\vec{x}_4) = \int d^d\vec{x}\, {}^{\left(\text{AdS}\right)}{\tilde {\cal A}}^\text{c}_{\Delta_1\Delta_2 \Delta}(\vec{x}_1,\vec{x}_2,\vec{x})\, {}^{\left(\text{AdS}\right)}{\tilde {\cal A}}^\text{c}_{\Delta_3\Delta_4 d-\Delta}(\vec{x}_3,\vec{x}_4,\vec{x}).
\end{equation}
This gives the following CPWE for the exchange \eqref{cspexch}:
\begin{equation}
    {\tilde {\cal A}}^{\text{exch.}}_{\Delta_1\Delta_2|m^2|\Delta_3\Delta_4}=\left(\tfrac{d}{2}-\Delta\right)\int^{\frac{d}{2}+i\infty}_{\frac{d}{2}-i\infty} \frac{d\Delta}{2\pi i}\, \frac{c^{+ + -}_{\Delta_1 \Delta_2 \Delta}c^{+ - -}_{\Delta \Delta_3 \Delta_4}}{c_{\Delta}}\,{}^{\left(\text{AdS}\right)}{\cal F}^{12,34}_{\Delta,J}.
\end{equation}
Note that meromorphicity of the spectral density in $\Delta$ then follows from meromorphicity in EAdS and of the coefficients \eqref{cseadstods} and \eqref{coeffnptfinal}, as consistent with single-valuedness of (perturbative) celestial correlators as a function of complex conformal cross ratios. This shows that in celestial CFT, which are not unitary in the familiar sense of Lorentzian CFTs and thus not necessarily endowed with a convergent OPE expansion, a convergent OPE expansion is recovered from the residues of the spectral density at any order in perturbation theory.

\paragraph{Unitarity.} In Euclidean CFT the conformal partial wave expansion \eqref{cpwe} follows by expanding in unitary representations of the Euclidean conformal group $SO\left(d+1,1\right)$. In particular, one expands into contributions from unitary irreducible representations of $SO\left(d+1,1\right)$ by inserting a resolution of the identity\footnote{In the following for simplicity we assume that only principal series representations contribute. In the most general case the above resolution of the identity would also involve discrete and complementary series representations of $SO\left(d+1,1\right)$.}
\begin{equation}
    {\mathds 1} = | \Omega \rangle \langle \Omega | + \sum^\infty_{J=0}\int^{\frac{d}{2}+i\infty}_{\frac{d}{2}}\frac{d\Delta}{2\pi i}\,\frac{1}{N\left(\Delta,J\right)}\int d^d\vec{x}\,| \Delta, J , \vec{x} \rangle \langle \Delta, J , \vec{x} |,
\end{equation}
where ${\cal P}_{\Delta,J}$ projects onto the conformal multiplet labelled by the scaling dimension $\Delta$ and spin $J$, and $N\left(\Delta,J\right)$ is related to the norm of the state. Taking now
\begin{equation}
    {\cal O}_1 =  {\cal O}^\dagger_3 \quad \text{and} \quad {\cal O}_2 =  {\cal O}^\dagger_4,
\end{equation}
it is straightforward to conclude the following non-perturbative positivity constraint on the spectral density:
\begin{equation}\label{possd}
    \rho_J\left(\Delta\right) \geq 0.
\end{equation}
This follows from positivity of $N\left(\Delta,J\right)>0$, which is ensured by the unitarity of Principal Series representations in Euclidean CFT.

\vskip 4pt
Positivity \eqref{possd} of the spectral density has also been observed in the context of the conformal partial wave expansion of boundary correlators in dS space \cite{Hogervorst:2021uvp,DiPietro:2021sjt}. Here we are noting that it applies more generally to \emph{any} unitary Euclidean CFT, and therefore also to celestial correlators. It would be interesting to use \eqref{possd} to derive non-perturbative constraints on bulk Minkowski physics along the lines proposed in \cite{Hogervorst:2021uvp} and compare with other implementations of unitarity.\footnote{Note that the positivity \eqref{possd} is the analogue of the positivity of conformal block coefficients in Lorentzian CFT (dual to AdS physics), which is essential in formulating the numerical conformal bootstrap \cite{Rattazzi:2008pe,Poland:2018epd}.}

\vskip 4pt
Let us finally note that the above discussion holds for canonically normalised operators. In the context of celestial CFTs, we have a two-fold degeneracy of operators with the same dimension, one for in-coming and one-for outgoing particles. Furthermore, their 2pt functions are not canonical:
\begin{subequations}
\begin{align}
    \left\langle\mathcal{O}_1^+(x_1)\mathcal{O}_2^+(x_2)\right\rangle&=0\,,\\
    \left\langle\mathcal{O}_1^-(x_1)\mathcal{O}_2^-(x_2)\right\rangle&=0\,,\\
    \left\langle\mathcal{O}_1^+(x_1)\mathcal{O}_2^-(x_2)\right\rangle&=\frac{\mathcal{N}_\Delta}{(x_{12}^2)^{\Delta}}\,.
\end{align}
\end{subequations}
A canonical spectral representation can only be obtained in terms of operators which are normalised canonically. This amounts to consider:
\begin{subequations}
\begin{align}
    \mathcal{O}_{>}(x)&=\frac{1}{\sqrt{2}}\left(\mathcal{O}_1^+(x)+\mathcal{O}_1^-(x)\right)\,,\\
    \mathcal{O}_{<}(x)&=\frac{1}{\sqrt{2}}\left(\mathcal{O}_1^+(x)-\mathcal{O}_1^-(x)\right)\,.
\end{align}
\end{subequations}
Positivity conditions can now be derived also for celestial CFTs working with $\mathcal{O}_{>/<}$.

\section*{Acknowledgments}

CS and MT thank Andrea Puhm, Shu-Heng Shao and Kostas Skenderis for useful discussions and correspondence. CS and MT thank the Simons foundation for hospitality during the Symposium on Amplitudes Meet Cosmology 2022 and LI the TFI 2022 workshop of the INFN, where this work was presented. The research of CS was partially supported by the STFC grant ST/T000708/1. The research of MT was partially supported by the program  “Rita  Levi  Montalcini”  of the MIUR (Minister for Instruction, University and Research) and the INFN initiative STEFI.

\appendix

\section{Kantorovich-Lebedev transform}
\label{sec::KLT}

As reviewed in section \ref{sec::CCasWD}, the conformal primary wave function for a massive particle involves a Bessel function in the radial direction \cite{Pasterski:2016qvg,Pasterski:2017kqt} which, which as noted in \cite{deBoer:2003vf}, is related to the Kantorovich-Lebedev transform. This transform generalises the Mellin transform that implements the change of basis from scattering amplitudes to Celestial correlators in the case of massless particles. 

\vskip 4pt
Let us first review an analogous problem in EAdS$_{d+1}$.\footnote{The same applies to dS$_{d+1}$ (see \cite{Sleight:2021plv}) but we stick to EAdS$_{d+1}$ for concreteness.} In directions with translation invariance it is natural consider a decomposition in exponential plane waves, which diagonalise the translation generator. Such a decomposition is obtained via Fourier transform. In EAdS we only have translation invariance in the boundary directions. Perpendicular to the boundary it is instead natural to expand in terms of Eigenvalues of the Dilatation generator \cite{Sleight:2021plv}. In Poincar\'e coordinates \eqref{poincoords} the direction perpendicular to the boundary is parameterised by the coordinate $z$ and the Dilatation generator is diagonalised by power-laws:
\begin{align}\label{powers}
  f_\alpha(z)= \langle z |f_\alpha\rangle=z^{-i\alpha+\tfrac{d}{2}}\,,
\end{align}
which satisfy completeness and orthogonality:
\begin{subequations}
\begin{align}
    \langle f_\alpha|f_\beta\rangle&=2\pi\delta(\beta-\alpha)\,,\label{powerOrto}\\
    \int_{-\infty}^{+\infty} \frac{d\alpha}{2\pi}\langle z_1|f_\alpha\rangle\langle f_\alpha|z_2\rangle&=z_1^{d+1}\delta(z_1-z_2)\,.\label{powerCompl}
\end{align}
\end{subequations}
Any element of $L^2(\mathbb{R}^+,\frac{dz}{z^{d+1}})$ can be expanded in terms of power laws \eqref{powers} and this is implemented by the Mellin transform. See \cite{Sleight:2021plv} for more details. 
 
\vskip 4pt
In the context of Celestial holography, an orthogonal basis to expand elements of $L^2(\mathbb{R}^+,dR\,R^{d-1})$ is given by Bessel-K functions:
\begin{align}
  \tilde{K}_\alpha(R)= \langle R |\tilde{K}_\alpha\rangle=\frac{2R^{-d/2}}{\Gamma(i\alpha)}\,{K}_{i\alpha}(m R)\,,\label{BesselMode}
\end{align}
which are complete and orthogonal:\footnote{Note that the integral over $\alpha$ can be restricted to the interval $[0,\infty)$ removing the factor of $1/2$ and dropping the $\delta(\alpha+\beta)$.}
\begin{subequations}
\begin{align}
    \langle \tilde{K}_\alpha|\tilde{K}_\beta\rangle&=2\pi\delta(\beta-\alpha)+\frac{2 \pi  \Gamma (i \alpha) \delta (\alpha+\beta)}{\Gamma (-i \alpha)}\,,\label{BesselOrto}\\
    \frac12\int_{-\infty}^{+\infty} \frac{d\alpha}{2\pi}\langle R_1|\tilde{K}_\alpha\rangle\langle \tilde{K}_\alpha|R_2\rangle&=R_1^{-d+1}\delta(R_1-R_2)\,.\label{BesselCompl}
\end{align}
\end{subequations}
The above completeness and orthogonality relations can be shown straightforwardly by employing the Mellin-Barnes representation of the Bessel function. To prove \eqref{BesselOrto} one first goes to Mellin space rewriting it as:
\begin{multline}
    \langle \tilde{K}_\alpha|\tilde{K}_\beta\rangle=\int_{\epsilon}^\infty \frac{dR}{R}\,\int_{-i\infty}^{+i\infty}\frac{ds dt}{(2\pi i)^2}\,R^{\kappa-2 s-2 t}\left(\frac{m}{2}\right)^{-2 s-2 t} \,\\ \times \frac{\Gamma (s-\tfrac{i \alpha}{2}) \Gamma (s+\tfrac{i \alpha}{2}) \Gamma (t-\tfrac{i \beta}{2}) \Gamma (t+\tfrac{i \beta}{2})}{\Gamma (-i \alpha) \Gamma (i \beta)},
\end{multline}
where we have introduced two cut-off $\epsilon$ and $\kappa$. At this point the radial integral can be replaced with:
\begin{align}
    \int_{\epsilon}^\infty dR\, R^{\alpha-2s-2t}&=\frac{\epsilon ^{\alpha -2 s-2 t}}{2 s+2 t-\kappa}\,,& \Re(2s+2t-\kappa)>0\,.
\end{align}
After closing the contour on the left according to the inequality above, only one term survives in the $\epsilon\to0$ limit. The remaining Mellin-Barnes integral can then be evaluated using Barnes' first lemma, giving:
\begin{align}
    \langle \tilde{K}_\alpha|\tilde{K}_\beta\rangle=\lim_{\kappa\to 0}\frac{2^{\alpha -1} m^{-\alpha } \Gamma \left(\frac{1}{2} (\kappa \pm i \alpha\pm i\beta))\right)}{\Gamma (\kappa ) \Gamma (-i \alpha) \Gamma (i \beta)}=2 \pi  \delta (\beta-\alpha)+\frac{2 \pi  \Gamma (i \alpha) \delta (\alpha+\beta)}{\Gamma (-i \alpha)}\,.
\end{align}
By taking both $\alpha$ and $\beta$ positive, as usually considered in this context, the second term can be dropped and \eqref{BesselOrto} follows. To prove \eqref{BesselCompl}, one can first show that:
\begin{align}\label{Bessel2Int}
    \int_{0}^\infty \frac{dR_1}{R_1}\underbrace{\int_{-\infty}^{+\infty}\frac{d\alpha}{2\pi}\frac{K_{i\alpha}(R_1)}{\Gamma(i\alpha)}\frac{K_{-i\alpha}(R_2)}{\Gamma(-i\alpha)}}_{\equiv\, k(R_1,R_2)}=\frac12\,,
\end{align}
from which it follows that\footnote{It is straightforward, picking up the residues in the $\alpha$ complex plane, to show that for $R_1\neq R_2$ the integral over $\alpha$ is vanishing.}
\begin{align}
    k(R_1,R_2)=\frac{R_1}{2}\delta(R_1-R_2)\,.
\end{align}
From this \eqref{BesselCompl} follows upon multiplying by the appropriate powers of $R$ and changing variables. To prove \eqref{Bessel2Int}, as we did above for the orthogonality relation one uses the Mellin-Barnes representation for the Bessel function: 
\begin{multline}
    \int_{0}^\infty \frac{dR_1}{R_1} k(R_1,R_2)=\int_{0}^\infty \frac{dR_1}{R_1}\int_{-\infty}^{+\infty}\frac{d\alpha}{2\pi}\int_{-i\infty}^{+i\infty}\frac{ds\,dt}{(2\pi i)^2}\,\left(\tfrac{R_1}2\right)^{-2 s} \left(\tfrac{R_2}2\right)^{-2 t}\,\\ \times \frac{\Gamma (s-\frac{i \alpha }{2}) \Gamma (s+\frac{i \alpha }{2}) \Gamma (t-\frac{i \alpha }{2}) \Gamma (t+\frac{i \alpha }{2})}{\Gamma (-i \alpha) \Gamma (i \alpha )}.
\end{multline}
One performs the $R_1$ integration by placing a cut-off around zero:
\begin{align}
    \lim_{\epsilon\to0} \int_{0}^\infty \frac{dR_1}{R_1}\,R_1^{-2s}&=\lim_{\epsilon\to0}\frac{\epsilon^{-2s}}{2s}\,,& \Re(s)&>0\,.
\end{align}
At this point one can close the $s$ contour of integration to the left. Picking the contribution of the $s\sim0$ residue one obtains:
\begin{align}
    \int_{0}^\infty \frac{dR_1}{R_1}\,k(R_1,R_2)=\int_{-i\infty}^{+i\infty}\frac{dt}{2\pi i}\int_{-\infty}^{+\infty}\frac{d\alpha}{2\pi}\,\cosh \left(\tfrac{\pi  \alpha }{2}\right) \Gamma (t-\tfrac{i \alpha }{2}) \Gamma (t+\tfrac{i \alpha }{2}).
\end{align}
One performs first the $\alpha$ and then the $t$ integral simply closing the contours and picking the residues:
\begin{align}
    \int_{0}^\infty \frac{dR_1}{R_1}\,k(R_1,R_2)=\sum_{n,m=0}^\infty\frac{(-1)^n\left(\frac{R_2}{2}\right)^{m+n} \cos \left(\frac{\pi}{2}   (m+n)\right)}{2n! m!}\,.
\end{align}
Upon changing summation variables as $\tilde{n}=n-m$ and one can perform the sum over $m$ using the identity:
\begin{align}
    \sum_{m=0}^{\infty}\frac{(-1)^{-m}}{\Gamma (m+1) (\tilde{n}-m)!}=\delta_{\tilde{n},0},
\end{align}
which collapses the leftover sum over $\tilde{n}$ to obtain \eqref{Bessel2Int}.

\vskip 4pt
The decomposition of an element of $L^2(\mathbb{R}^+,dR\,R^{d-1})$ in terms of functions \eqref{BesselMode} is implemented by the Kantorovich-Lebedev Transform. For a field $\phi\left(X\right)$ in $\mathbb{M}^{d+2}$, this provides a map to fields $\phi_\Delta(\hat{X})$ living the $H^+_{d+1}$ hypersurface and vice versa:
\begin{subequations}
\begin{align}
    \phi(R,\hat{X})&=\frac12\int_{-\infty}^{+\infty} \frac{d\nu}{2\pi}\,\phi_{\frac{d}{2}+i\nu}(\hat{X})\,\tilde{K}_\nu(R)\,,\\
    \phi_{\frac{d}{2}+i\nu}(\hat{X})&=\int_0^\infty dR\, \phi(R,\hat{X})\,\tilde{K}_\nu^*(R)\,.
\end{align}
\end{subequations}

\section{The conformal primary wave function}

\label{sec::CPWint}

In this Appendix we evaluate the integral 
\begin{align}\label{varphiA}
    \varphi_{\Delta}(X;Q)=\frac{\mathcal{N}_\Delta}{\Gamma(\Delta-\tfrac{d}{2})}\left(\frac{m}{2\pi}\right)^{d/2}\int_{H^+_{d+1}}[d{\hat p}]\,G_{\Delta}\left({\hat p};Q\right)\,e^{ + m \,{\hat p}\cdot X}\,, \quad X \in {\cal A}_+,
\end{align}
This integral is convergent for $X \in {\cal A}_+$ and is the starting point to define the analytic continuation of the conformal primary wave functions over all Minkowski space.

\vskip 4pt
To evaluate the above integral we apply Schwinger representation for the EAdS bulk-boundary propagator \eqref{buboEAdS}:
\begin{align}\label{buboschw}
    G_\Delta(\hat{p};Q)=\frac{C^{\text{AdS}}_{\Delta}}{\Gamma(\Delta)}\int_0^\infty\frac{dt}{t}\,t^\Delta e^{2t\hat{p}\cdot Q}\,.
\end{align}
Plugging into \eqref{varphiA} gives
\begin{align}
    \varphi_{\Delta}(X;Q)=\frac{\mathcal{N}_\Delta}{\Gamma(\Delta-\tfrac{d}{2})}\left(\frac{m}{2\pi}\right)^{d/2}\frac{C^{\text{AdS}}_{\Delta}}{\Gamma(\Delta)}\int_{H^+_{d+1}}[d{\hat p}]\,\int_0^\infty\frac{dt}{t}\,t^\Delta \,e^{2\,{\hat p}\cdot T}\,,
\end{align}
where, in the hyperbolic slicing $X = R {\hat X}$, we introduced the time-like vector
\begin{equation}
  T = t Q+\frac12 m R {\hat X},
\end{equation}
which can be written in the form
\begin{equation}
    T = |T|(1,1,\vec{0}).
\end{equation}
To integral over $H^+_{d+1}$ can then be evaluated by employing the parameterisation:
\begin{align}
    \hat{p}=\frac1{x_0}(1,x_0^2+\vec{x}^2,\vec{x})\,.
\end{align}
Evaluating the Gaussian integral over $\vec{x}$ gives
\begin{align}
    \varphi_\Delta(X;Q)=\frac{\mathcal{N}_\Delta}{\Gamma(\Delta-\tfrac{d}{2})}\left(\frac{m}{2}\right)^{d/2}\frac{C^{\text{AdS}}_{\Delta}}{\Gamma(\Delta)}\int_{0}^\infty\frac{dt}{t}\,t^{\Delta}\int_{0}^\infty\frac{dx_0}{x_0}\,x_0^{-d/2}\,e^{-x_0+\frac{T^2}{x_0}}\,,
\end{align}
where
\begin{align}
    T^2=-\frac{m^2 R^2}{4}+m R t{\hat X}\cdot Q\,.
\end{align}
The integral over $t$ can then replaced with the bulk-to-boundary propagator \eqref{buboschw}:
\begin{equation}
    \varphi_\Delta(X;Q)=R^{-d/2}\frac{\mathcal{N}_\Delta}{\Gamma(\Delta-\tfrac{d}{2})} \left(\frac{2}{mR}\right)^{\Delta-\frac{d}{2}} G_{\Delta}({\hat X};Q)\int^\infty_0 \frac{dx_0}{x_0} x^{\Delta-\frac{d}{2}}_0 e^{-x_0}e^{-\frac{m^2R^2}{4x_0}}.
\end{equation}
The remaining integral over $x_0$ can be performed using the Mellin-Barnes representation for the exponential function. In particular,
\begin{subequations}
\begin{align}
    e^{-x_0} &= \int^{+i\infty}_{-i\infty}\frac{ds_1}{2\pi i}\,\Gamma\left(s_1\right) x^{-s_1}_0,\\
    e^{-\frac{m^2R^2}{4x_0}} &= \int^{+i\infty}_{-i\infty}\frac{ds_2}{2\pi i}\,\Gamma\left(s_2\right)\left(\frac{m^2R^2}{4}\right)^{-s_2} x^{s_2}_0.
\end{align}
\end{subequations}
The integral over $x_0$ can then be evaluated to a Dirac delta function:
\begin{align}
     \int^{\infty}_0 \frac{dx_0}{x_0}\,x^{\Delta-\frac{d}{2}+s_2-s_1}_0=2\pi i\,\delta\left(s_1-s_2-\Delta+\tfrac{d}{2}\right).
\end{align}
This can be used to eliminate one of the two Mellin-Barnes integrals, obtaining: 
\begin{align}\nonumber
    \varphi_\Delta(X;Q)&=\frac{R^{-d/2}\mathcal{N}_\Delta}{\Gamma(\Delta-\tfrac{d}{2})}G_{\Delta}({\hat X}_{\text{AdS}};Q)\,\int_{-i\infty}^{+i\infty}\frac{ds}{2\pi i}\,\Gamma\left(s-\tfrac12\left(\Delta-\tfrac{d}{2}\right)\right)\Gamma\left(s+\tfrac12\left(\Delta-\tfrac{d}{2}\right)\right)\left(\frac{m R}2\right)^{-2s}\,,\\
    &=\mathcal{N}_\Delta\, G_{\Delta}({\hat X}_{\text{AdS}};Q) \tilde{K}_{\Delta-\tfrac{d}2}(m R),
\end{align}
where in the last equality we identified the Mellin-Barnes representation \eqref{MBrepKtilde} of $\tilde{K}_{\Delta-\tfrac{d}2}(m R)$.

\vskip 4pt
Note that this demonstrates how standard embedding space techniques (see e.g. \cite{Penedones:2010ue,Paulos:2011ie,Fitzpatrick:2011ia}) for evaluating Witten diagrams in EAdS can be adapted to Celestial correlators.

\bibliographystyle{JHEP}
\bibliography{refs}

\end{document}